\documentclass[manuscript]{aastex}

\usepackage{pstricks,pst-node,pst-tree}

\usepackage{amsmath}

\newcommand{\E}[1]{\times 10^{#1}}   
\newcommand{\EE}[1]{$\times 10^{#1}$}   
\newcommand{\reac}[6]{$^{#1}{\rm #2}(#3,#4)^{#5}{\rm #6}$}   
\newcommand{\xx}[2]{$^{#1}{\rm #2}$} 
\newcommand{\xxf}[2]{^{#1}{\rm #2}} 
\newcommand{\nn}[2]{^{#1}{\rm #2}} 
\newcommand{\YY}[2]{Y({^{#1}{\rm #2}})} 
\newcommand{\YYSQ}[2]{Y^2({^{#1}{\rm #2}})} 
\newcommand{\YYb}[2]{\bar Y({^{#1}{\rm #2}})} 
\newcommand{\YYt}[2]{\tilde Y({^{#1}{\rm #2}})} 
\newcommand{\rr}[1]{\rho N_A \lambda_{#1}}
\slugcomment{Not to appear in Nonlearned J., 45.}

\shorttitle{Simplified Hydrostatic C--burning in WD interiors.}
\shortauthors{F\"orster, Lesaffre \& Podsiadlowski}

\begin{document}

\title{Simplified Hydrostatic Carbon Burning in White Dwarf
  Interiors.}  \author{Francisco F\"orster\altaffilmark{1,3} Pierre
  Lesaffre\altaffilmark{2} and Philipp Podsiadlowski\altaffilmark{3}}
\altaffiltext{1}{Departamento de Astronom\'ia, Universidad de Chile,
  Casilla 36-D, Santiago, Chile} \altaffiltext{2}{Laboratoire de
  Radioastronomie, 24 rue Lhomond, 75231 PARIS Cedex 05, France}
\altaffiltext{3}{University of Oxford, Department of Physics, Oxford,
  OX1 3RH, UK}

\begin{abstract}
We introduce two simplified nuclear networks that can be used in
hydrostatic carbon burning reactions occurring in white dwarf
interiors. They model the relevant nuclear reactions in carbon--oxygen
white dwarfs (CO WDs) approaching ignition in Type Ia supernova (SN
Ia) progenitors, including the effects of the main $e^-$--captures and
$\beta$--decays that drive the convective Urca process. They are based
on studies of a detailed nuclear network compiled by the authors and
are defined by approximate sets of differential equations whose
derivations are included in the text. The first network, N1, provides a
good first order estimation of the distribution of ashes and it also
provides a simple picture of the main reactions occuring during this
phase of evolution. The second network, N2, is a more refined version
of N1 and can reproduce the evolution of the main physical properties
of the full network to the 5\% level. We compare the evolution of the
mole fraction of the relevant nuclei, the neutron excess, the photon
energy generation and the neutrino losses between both simplified
networks and the detailed reaction network in a fixed temperature and
density parcel of gas.

\end{abstract}


\section{INTRODUCTION}
Type Ia supernovae (SNe Ia) are the thermonuclear explosions of white
dwarf stars. They are observable end--points of stellar evolution,
they shape the energy and chemistry evolution of galaxies and have
been successfully used as distance indicators up to redshifts of $\sim
1.7$ thanks to an empirical decline rate -- luminosity relation
\citep{1977SvA....21..675P, 1993ApJ...413L.105P}. This relationship
led to the discovery of the acceleration of the Universe
\citep{1998AJ....116.1009R, 1999ApJ...517..565P}

The exact nature of SN Ia progenitors is still debated
\citep{2000ARA&A..38..191H}, but most models involve an accreting
carbon--oxygen white dwarf (CO WDs) with a mass close to the
Chandrasekhar mass \citep{1984ApJ...286..644N}. This CO WD would
accrete mass from either a slightly evolved main sequence companion
(CO WD + MS), a more evolved red giant star (CO WD + RG), or a Helium
star (CO WD + He), in the so--called \emph{single degenerate}
scenarios \citep[\emph{SD}, see e.g.][]{1996ApJ...470L..97H,
  1997A&A...322L...9L, 1999ApJ...522..487H, 1999ApJ...519..314H,
  2000A&A...362.1046L, 2004MNRAS.350.1301H, 2009MNRAS.395.2103M}, or
as a result of the merger of two degenerate stars with a combined mass
above the Chandrasekhar limit, in the \emph{double degenerate}
scenario \citep[\emph{DD}, see e.g.][]{1984ApJ...277..355W,
  1984ApJS...54..335I}. It has been suggested that only CO WDs
accreting within a narrow range lead to successful thermonuclear
explosions \citep[see e.g.][]{1991ApJ...367L..19N,
  2007ApJ...663.1269N}, unless additional physical processes are
included in the models, e.g. rotation and other three--dimensional
effects \citep[see e.g.][]{2006ApJ...644...21D, 2007MNRAS.380..933Y,
  2010Natur.463...61P}.

The diversity of SN Ia ejecta and the origin of the decline
rate--absolute magnitude relation are being understood only recently
thanks to new observational techniques and energy transfer codes
\citep[see e.g.][]{2007Sci...315..825M, 2007ApJ...662..487W,
  2009MNRAS.398.1809K}. These efforts have been accompanied by new
developments in the physics of the explosion which have shown that
even if pure deflagration models can reproduce many SN Ia spectra and
light curves \citep{2007ApJ...668.1132R}, in some cases a delayed
detonation may be necessary \citep{1991A&A...245..114K,
  2007ApJ...668.1132R}. The physics of both the transition to
detonation and ignition of the deflagration wave remain uncertain
\citep{2006A&A...450..655I, 2007ApJ...668.1103R, 2010arXiv1001.2165I}.

A related problem, rarely addressed in the literature, is how to
connect theoretical models with observed systematic differences
between SNe Ia occurring in different stellar environments
\citep[e.g.][]{1996AJ....112.2391H, 2006ApJ...648..868S}. These should
be related to the pre--supernova evolution and not to line--of--sight
effects or other processes occuring randomly during the explosion
\citep[e.g.][]{2009Natur.460..869K, 2010Natur.466...82M}. Thus,
presupernova evolution must play a significant role in the diversity
of SN Ia explosions \citep{2006MNRAS.368..187L}.

\subsection{Presupernova Evolution: from Cooling to Ignition}
Here we will assume that the progenitors of SNe Ia originate in the SD
scenario; but note that, even in the standard DD scenario, the core
would evolve in a very similar way in the immediate pre-explosion
phase \citep[i.e. during the last $\sim 10^3\,$yr;
  see][]{2007MNRAS.380..933Y}.

Before a SN Ia progenitor becomes unbound by the explosion it
undergoes several distinct phases of evolution
\citep{1984ApJ...286..644N, 2006MNRAS.368..187L}. First, the
progenitor white dwarf cools down at almost constant density after its
birth, for a period of hundreds or thousands of Myr, depending on the
particular formation scenario (the \emph{cooling phase}). Then, it
accretes matter for a period of the order of one Myr, making the
degenerate star shrink to keep the hydrostatic equilibrium and its
core compress adiabatically, the \emph{accretion phase}. Adiabatic
compression at the center and heat diffusion from the hot accreting
envelope can make the central temperature increase under degenerate
conditions, triggering hydrostatic carbon burning and a new source of
energy generation. The energy input from hydrostatic carbon burning
will force the star to transport the excess energy at its center
convectively, the \emph{simmering phase}. During this phase the
convective core can grow to engulf most of the star. If the central
density is high enough, $e^-$--captures and $\beta$--decays can become
important.  In the presence of a convective core, that process has
been called the \emph{convective Urca process}.

At some point energy deposition will dominate over the energy losses,
making the star's central temperature increase at almost constant
density, the \emph{thermonuclear flash}. Finally, when the temperature
is high enough, one or more ignition spots will give rise to nuclear
flames that will propagate in the highly convective medium, the
\emph{thermonuclear runaway}, causing a deflagration wave to sweep
over the star, sometimes transitioning into a detonation wave. The
deflagration and detonation waves will generate most of the kinetic
energy and ashes in the ejecta in a few seconds, including radioactive
matter which will later power the light curve of the supernova. Since
this last phase will occur at very high temperatures, the
characteristic burning time--scales will be much shorter than the
typical weak interaction time--scales and the neutron excess of the
ejecta will not differ from that of the presupernova star, except for
the star's innermost regions where weak interaction time--scales are
shorter. Most of the WD's neutron excess changes will occur before
ignition.

\subsection{Nuclear physics and the convective URCA process}

One of the main obstacles that remain to be solved in order to obtain
self--consistent pre--supernova models up to ignition is the
\emph{convective Urca process}, which was mentioned in the previous
section. The following factors make this phase of evolution difficult
to solve: (1) the appearance of a fast--growing convective core with a
very steep luminosity and composition gradient at its outer edge, the
so--called C--flash; (2) a high--density medium with $e^-$--capture
and $\beta$--decay time--scales similar to the convective
time--scales, which have an uncertain effect over the energy budget of
the star; (3) a steep density gradient which changes rapidly when the
WD approaches the Chandrasekhar mass; (4) a steep composition gradient
of the Urca pair \xx{23}{Na}--\xx{23}{Ne} at the threshold density for
electron captures, which moves inwards as the central density
increases and (5) an increasingly complex set of nuclear reactions as
carbon burns and pollutes the WD with its ashes, with a time--scale
that can be similar to $e^-$--capture and $\beta$--decay time--scales

It is not clear whether $e^-$--captures and $\beta$--decays of Urca
matter around a threshold density in a convective medium have a
cooling or heating effect over time. Many studies over the years have
reached different conclusions regarding this
point. \citet{1972ApL....11...53P} first suggested that Urca processes
have a stabilizing effect over carbon burning, leading to the
formation of a neutron star instead of a thermonuclear
explosion. \citet{1973ApJ...183L.125B} realized that $e^-$--captures
can cause heating by creating holes in the Fermi sea, which dominate
over the neutrino losses for the most important Urca
pairs. \citet{1974ApJ...194..537C} found that significant work must be
done to mantain convection when Urca processes occur, with a net
cooling effect, but \citet{1975Ap&SS..37..143R} pointed out that
convection cannot develop fast enough to prevent
heating. \citet{1990ApJ...355..602B} summarized the factors
controlling the Urca process, but it was later shown that the role of
the kinetic energy flux should have been included in their analysis
\citep{1996A&A...311..152M, 1999ApJ...523..381S}. More recently,
\citet{2005MNRAS.356..131L} showed that the heating effect of the Urca
process depends on the state of mixing of the convective core, that
the convective velocities are reduced by Urca processes and that
time--dependent computations with a full nuclear network are needed to
understand the effect of Urca processes on the ignition conditions of
SNe Ia.

In this study we have focused on how to accurately treat the
increasingly complex nuclear reactions as the WD is polluted with
ashes. We do not attempt to answer whether Urca processes cause
cooling or heating, but provide a tool for better evaluating the
competing heating and cooling mechanisms in a presupernova WD
approaching ignition.

In what follows we will introduce two approximations that use a
limited number of nuclei to accurately describe the evolution of the
main species that result from the burning of \xx{12}{C}, the energy
deposition rate, the energy losses via neutrinos and the neutron
excess. We will show when these approximations hold, and their
potential applications, but first we will introduce a detailed nuclear
network which will be used for comparison with the simplified
networks.

\section{THE FULL NUCLEAR NETWORK} \label{sec:full}

\begin{figure}[ht!]
  \centering
  \includegraphics[width=0.7\hsize,angle=270,keepaspectratio]{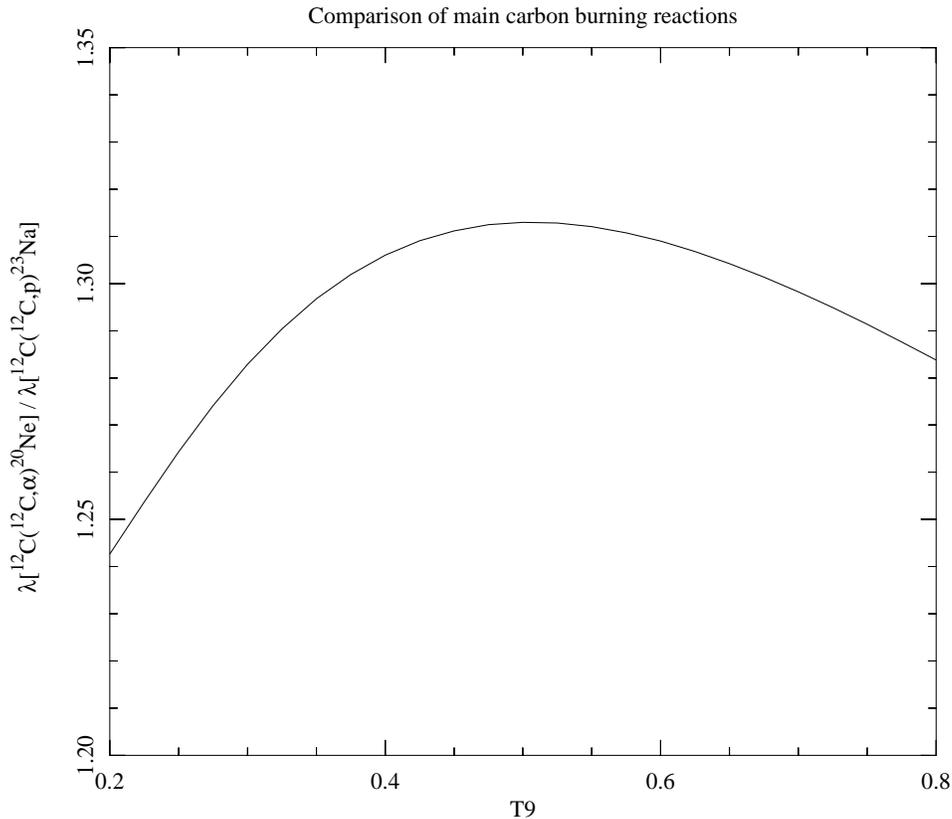}
  \caption[Main \xx{12}{C} burning reactions]{Ratio between the two
    main carbon burning reactions (thermally averaged cross--sections)
    plotted against the temperature in $10^9$ K.}
  \label{fig:comparison_12C}
\end{figure}

The nuclear reactions within a CO WD approaching explosion are
characterized by the burning of \xx{12}{C} nuclei at high densities
($>10^7$ g cm$^{-3}$) and high temperatures ($>10^8$ K) in an
environment rich in \xx{12}{C} and \xx{16}{O} nuclei and relatively
devoid of free protons, $\alpha$--particles or neutrons. The dominant
reactions are $\rm{^{12}C(^{12}C, p)^{23}Na}$, Q = 4.6 MeV, and
$\rm{^{12}C(^{12}C, ^{4}He)^{20}Ne}$, Q = 2.2 MeV, both occurring at
similar rates (see Figure~\ref{fig:comparison_12C}). The nuclear
network increases in complexity as the \xx{12}{C}--burning pollutes
the WD with its ashes, mainly \xx{20}{Ne} and \xx{23}{Na}, but also
with protons and $\alpha$--particles, which will be processed into
additional \xx{16}{O} and \xx{13}{C} nuclei, as will be shown later.

We have integrated a detailed reaction network at fixed temperature
and density compiled by the authors. The system of differential
equations defining the network is solved for using the semi--implicit
extrapolation method from \citet{baderdeuflhard83}. We include species
that are part of the reactions known to be most important, or those
close to them in a table of nuclides. We include all reaction rates
between species of our network that are either in the Reaclib library
\citep{reaclib}, or in the weak interaction tables by \citet{odaetal},
as well as improved \reac{13}{N}{e^-}{\nu_e}{13}{C} rates from
\citet{2008PhRvC..77b4307Z}. Screening corrections were also included
under the simplification of \citet{1973ApJ...181..457G}. Recent
formalisms that treat carbon burning and screening corrections in
alternative ways were not implemented for this work \citep[see
  e.g.][]{2005PhRvC..72b5806G, 2006PhRvC..74c5803Y,
  2007PhRvL..98l2501S}.

We have assumed that the initial C/O ratio is always
given by the nuclide mass fractions X(\xx{12}{C}) = 0.3 and X(\xx{16}{O}) =
0.7, in order to compare with the work by \citet{2008ApJ...677..160C},
but reasonable variations of the initial compositions do not affect
the validity of our approximations. 
\begin{figure*}
  \centering
  \includegraphics[width=0.65\hsize,angle=270,keepaspectratio]{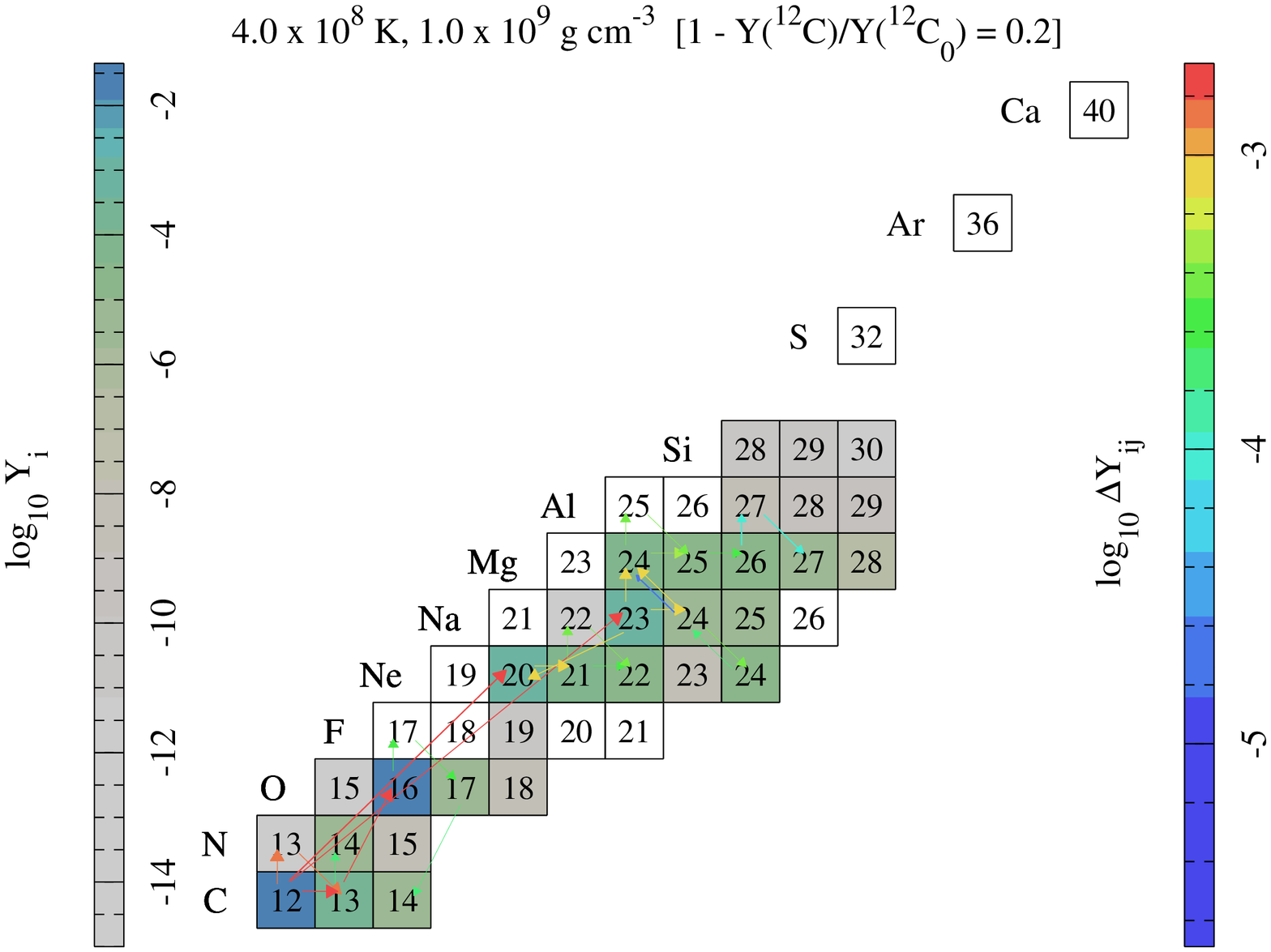}
  \includegraphics[width=0.65\hsize,angle=270,keepaspectratio]{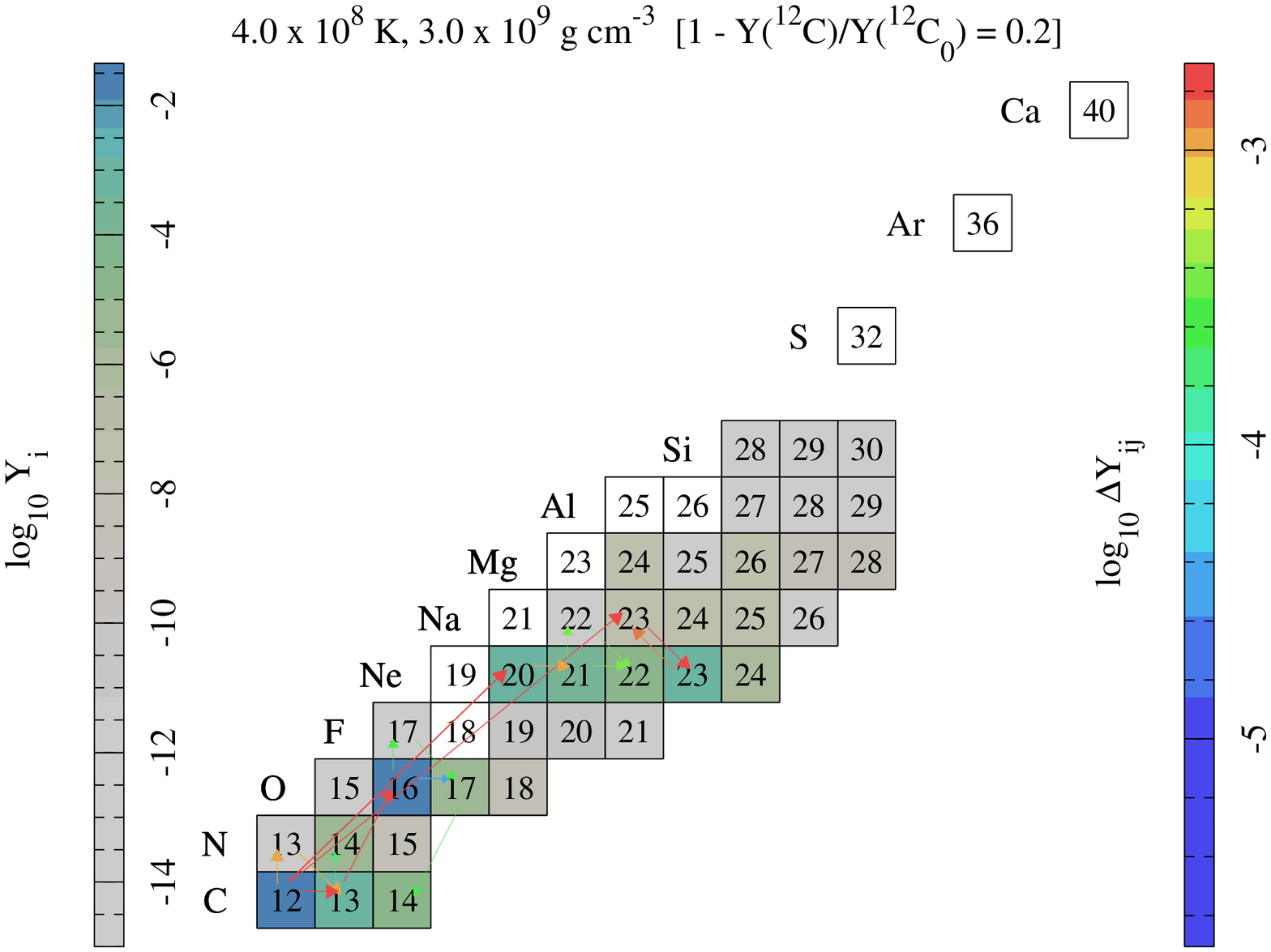}
  \caption[Flows at 3$\E{8}$ K in pure CO mixture up to 20\% burnt
    \xx{12}{C}.]{Main flows at a temperature of 4$\E{8}$K and
    densities of 1$\E{9}$ g cm$^{-3}$ (top) and 3$\E{9}$ g cm$^{-3}$
    (bottom) when 20\% of the original carbon has been burnt.}
  \label{fig:chap04_flows3e8_0p04_CO}
\end{figure*}
Figure~\ref{fig:chap04_flows3e8_0p04_CO} show the main flows in our
nuclear network at a temperature of $3 \E{8}$ K and densities of $1
\E{9}$ g cm$^{-3}$ (top) and $3 \E{9}$ g cm$^{-3}$ (bottom), starting
from a pure carbon--oxygen mixture, when 20\% of the original carbon
nuclei have been burnt. The left hand side color--coding bar indicates
the mole fraction scale of the individual species, whereas the right
hand side color--coding bar indicates the mole fraction flow scale of
the individual reactions. Only reactions that are bigger than one
thousandth of the biggest flow are plotted.

The main differences between the low density (top) and high density
(bottom) flow diagrams are due to $e^-$--captures in the
$^{23}$Na(e$^-$, $\nu_{e}$)$^{23}$Ne reaction, which only occurs at
densities higher than $\sim$1.7$\E{9}$ g cm$^{-3}$. Below this density
the $e^-$--captures are almost exclusively due to the
\reac{13}{N}{e^-}{\nu_e}{13}{C} reaction.

The advantage of this detailed network is that it can accurately track
the evolution of nuclei with a wide range of characteristic
time--scales. For example, it is capable of accurately following the
mole fraction of $\alpha$--particles, protons, neutrons or fast
electron--capturing \xx{13}{N} nuclei, as well as long characteristic
time--scale nuclei like \xx{12}{C}, or intermediate time--scale nuclei
like \xx{23}{Na} at high densities.

Since we study this reaction network in convective WD interiors, we
must consider the relation between the relevant burning time--scales
and the convective turn--over time--scale, which is determined by the
diffusion time--scale and the temperature, density and composition
gradients of the star. The convective time--scale will be the shortest
time--scale that affects the evolution of the structure of the star
that is not directly connected to the nuclear physics, and can be as
low as 50 sec before ignition \citep[see Figure~6
  in][]{2006MNRAS.368..187L}.

Ideally, we would like to build an approximated nuclear network where
we remove time--scales much shorter than the convective time--scale
from the resulting set of differential equations, as long as a
reasonable physical justification is provided. This would reduce the
number of variables per zone while keeping accuracy and would make the
inversion of the corresponding Jacobian in a Newton--Raphson
integration scheme more stable.

We have found two approximate solutions that achieve the former
based on the detailed nuclear network at fixed density and temperature
described in this section \citep[see also][]{2008ApJ...677..160C,
  2008ApJ...673.1009P}. The first such approximation will be described
in the following section.

\clearpage

\section{THE SIMPLIFIED NETWORK: FIRST APPROXIMATION (N1)} \label{sec:1stapprox}

To accurately follow the chemistry of the evolving WD progenitor
towards explosion, the simultaneous solution of the structure and
chemistry of the star is required \citep{2006MNRAS.370.1817S}. The
chemistry equations must be able to reproduce the evolution of the
main species and their effect on the star through changes in energy
deposition, energy losses or the electron fraction. This can become
computationally demanding if too many species are included and can
worsen during the \emph{simmering phase} of evolution due to known
convergence problems \citep{1978ApJ...226..996I, 1982ApJ...253..248I},
even with approximate theories of convection.

\begin{figure*}[ht!]
  \centering
  \includegraphics[width=0.7\hsize,keepaspectratio]{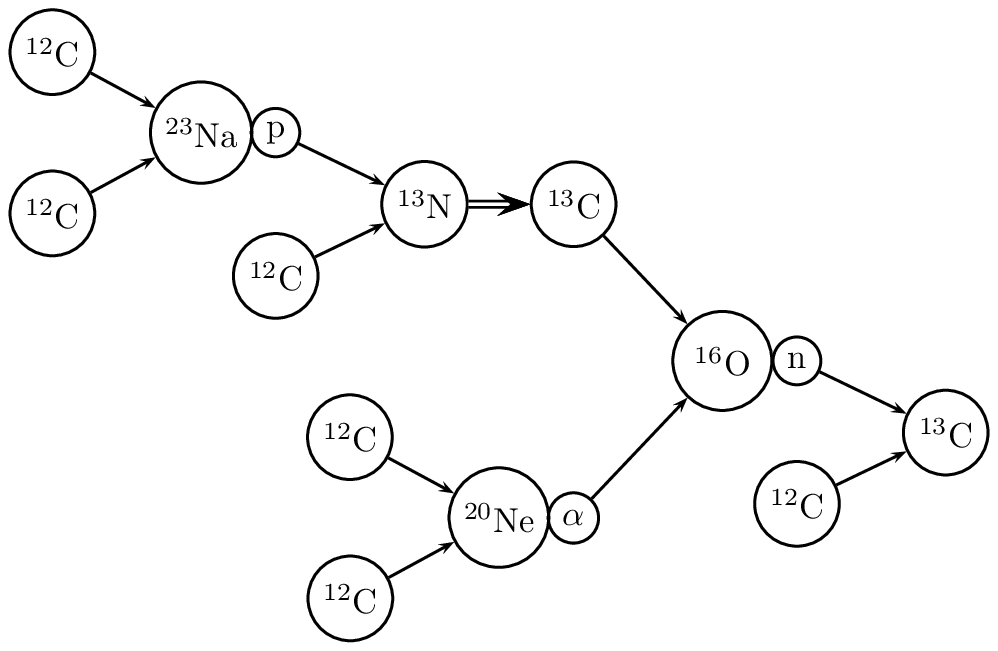}
         \caption[Low density \xx{12}{C} burning with proton
           leaks]{Main \xx{12}{C} burning reactions and its ashes at
           low density. Six \xx{12}{C} nuclei are consumed to produce
           four nuclei: \xx{23}{Na}, \xx{20}{Ne}, \xx{16}{O} and
           \xx{13}{C} and one electron capture in the reaction
           \reac{13}{N}{e^-}{\nu_e}{13}{C}. The figure is shown as a
           rotated tree, with parent nuclei connected to children
           nuclei in the direction indicated by the arrows. Double
           arrows correspond to electron captures or inverse beta
           decays.}
         \label{fig:treeC12low}
\end{figure*}

\begin{figure*}[ht!]
  \centering
  \includegraphics[width=0.7\hsize,keepaspectratio]{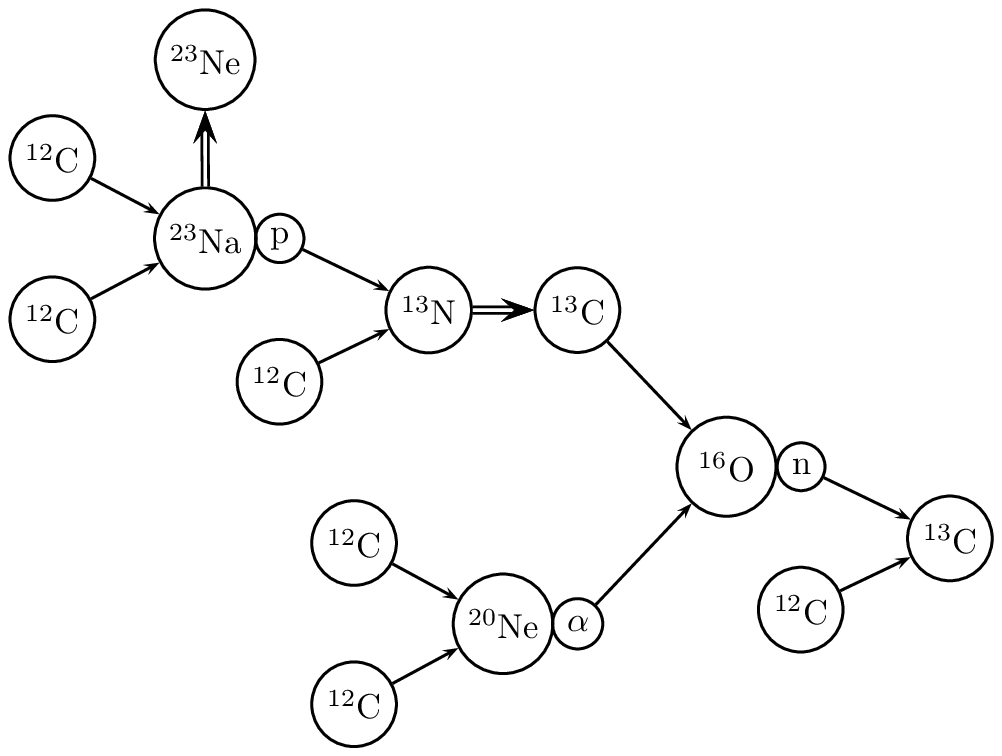}
         \caption[High density \xx{12}{C} burning with proton
           leaks]{Same as Figure~\ref{fig:treeC12low}, but at high
           density.  Six \xx{12}{C} nuclei are consumed to produce
           four nuclei: \xx{23}{Ne}, \xx{20}{Ne}, \xx{16}{O} and
           \xx{13}{C} and two $e^-$--captures in the reactions
           \reac{13}{N}{e^-}{\nu_e}{13}{C} and
           \reac{23}{Na}{e^-}{\nu_e}{23}{Ne}.}
         \label{fig:treeC12high}
\end{figure*}

We have found a first order approximation to the full nuclear network
which will be described in this section. We will refer to this
approximation as N1. A more accurate version of this approximation,
N2, will be described in Section~\ref{sec:2ndapprox}. Both simplified
networks use the fact that the dominant carbon burning reactions,
\reac{12}{C}{^{12}{\rm C}}{p}{23}{Na} and \reac{12}{C}{^{12}{\rm
    C}}{\alpha}{20}{Ne}, occur nearly at the same rate and the
abundances of protons, $\alpha$--particles, neutrons and \xx{13}{N}
nuclei are approximately at equilibrium most of the time. This can be
explained by the complementarity of the relevant reactions (see
Figure~\ref{fig:treeC12low} and \ref{fig:treeC12high}), which was
first noticed by \citet{2008ApJ...677..160C} and
\citet{2008ApJ...673.1009P}.

First, when in the carbon burning reaction \reac{12}{C}{^{12}{\rm
    C}}{p}{23}{Na} a proton is released, it will be captured quickly,
preferentially in the \reac{12}{C}{p}{\gamma}{13}{N} reaction. Second,
\xx{13}{N} will quickly capture an electron in the
\reac{13}{N}{e^-}{\nu_e}{13}{C} reaction, decreasing the pressure and
lifting the degeneracy of the gas, and producing a \xx{13}{C}
nucleus. Third, in the carbon burning reaction \reac{12}{C}{^{12}{\rm
    C}}{\alpha}{20}{Ne} an $\alpha$--particle will be released and
quickly captured in the \reac{13}{C}{\alpha}{n}{16}{O} reaction, thus
consuming the \xx{13}{C} nucleus just produced, but liberating a
neutron. Fourth, the free neutron will be preferentially captured in
the \reac{12}{C}{n}{\gamma}{13}{C} reaction, thus recovering the
\xx{13}{C} just consumed. Hence, the net effect is approximately the
burning of six \xx{12}{C} nuclei and the capture of one electron to be
replaced by four nuclei: \xx{20}{Ne}, \xx{23}{Na}, \xx{16}{O} and
\xx{13}{C}, i.e.:
\begin{align}
  6~\nn{12}{C} + e^- ~\rightarrow~ \nn{23}{Na}~ +~ \nn{20}{Ne}~ +~
  \nn{16}{O}~ +~ \nn{13}{C}. 
\end{align}
An additional $e^-$--capture can occur via the
\reac{23}{Na}{e^-}{\nu_e}{23}{Ne} reaction if the density is above
$\approx 1.7\E{9}$ g cm$^{-3}$. In this case, two $e^-$--captures can occur
and the net effect is:
\begin{align}
  6~\nn{12}{C} + 2~e^- ~\rightarrow~ \nn{23}{Ne}~ +~ \nn{20}{Ne}~ +~
  \nn{16}{O}~ +~ \nn{13}{C},
\end{align}
which is schematically shown in Figure~\ref{fig:treeC12high}. If any
of the flows above is broken the simplified network will fail and will
need a different treatment.

Now, we will assume that the only relevant reactions are:

\begin{tabular}{@{}l l l l l l @{}}
1)&  \reac{12}{C}{^{12}{\rm C}}{p}{23}{Na}, & Q = 2.24 MeV, & 2)&  \reac{12}{C}{^{12}{\rm C}}{\alpha}{20}{Ne}, & Q = 4.62 MeV \\
3)&  \reac{12}{C}{p}{\gamma}{13}{N}, & Q = 1.94 MeV, & 4)&  \reac{13}{N}{e^-}{\nu_e}{13}{C},            & Q = 2.22 MeV \\
5)&  \reac{13}{C}{\alpha}{n}{16}{O}, & Q = 2.22 MeV, & 6)&  \reac{12}{C}{n}{\gamma}{13}{C},             & Q = 4.95 MeV \\
7)&  \reac{23}{Na}{e^-}{\nu_e}{23}{Ne}, & Q = -4.38 MeV, & 8)&  $^{23}{\rm Ne}(\beta-)^{23}{\rm Na}$,       &  Q = 4.38 MeV. \\
\end{tabular}

\vspace{.1cm} Note that only mass differences are used to compute the
Q--values and not the energy of the electrons lost or gained in weak
interactions and that the \xx{13}{N} positron decay reaction must
also be included at low densities. The change in electron density,
which can be computed assuming charge conservation, can be used with
the chemical potential of the electrons to compute the energy changes
due to electron gains or losses. The Fermi energy of the electrons at
high densities is approximately $5.1$ MeV $[ \rho Y_{\rm e} / 10^9$ g
  cm$^{-3} ]^{1/3}$ \citep{2008ApJ...677..160C}.

Since ignition can occur at temperatures as high as $8\E{8}$ K
\citep[see Figure 5 in][]{2006MNRAS.368..187L}, it is necessary to
consider whether any inverse reaction from the list above can be
significant. We have also found that only the inverse of reaction 3),
i.e.  \reac{13}{C}{\gamma}{p}{12}{C}, needs to be considered, since
its characteristic time--scale can be comparable or smaller than the
characteristic time--scale of \xx{13}{N} e$^-$--captures, e.g. we have
found that both time--scales are similar for a temperature of $8\E{8}$
K and a density of $4\E{8}$ g cm$^{-3}$. Note that we use reaction
rates for the \reac{13}{N}{e^-}{\nu_e}{13}{C} reaction from
\citet{2008PhRvC..77b4307Z}.

Hence, we consider the following species: \xx{12}{C}, \xx{13}{C},
\xx{13}{N}, \xx{16}{O}, \xx{20}{Ne}, \xx{23}{Na}, \xx{23}{Ne}, $p$,
$\alpha$ and $n$.  Defining $\lambda_i \equiv \langle \sigma v
\rangle_i$ as the thermally averaged cross--section for the strong
interactions 1 to 6, or the rate of occurrence per particle per unit
time per unit volume of the weak interactions 7 and 8, we can write
the following differential equations:
\begin{align}
 \label{eq:C12}
d\YY{12}{C} / dt &= -\YYSQ{12}{C} \rr{1} - \YYSQ{12}{C} \rr{2} \notag \\ &-
\YY{12}{C} Y(p) \rr{3} - \YY{12}{C} Y(n) \rr{6} + \YY{13}{N} \lambda_3^{\rm inv},\\
\label{eq:C13}
d\YY{13}{C} /dt &= \YY{12}{C} Y(n) \rr{6} - \YY{13}{C} Y(\alpha)
\rr{5} + \YY{13}{N} \lambda_4,\\
\label{eq:13N}
d\YY{13}{N} /dt &= \YY{12}{C} Y(p) \rr{3} - \YY{13}{N} \lambda_4 - \YY{13}{N} \lambda_3^{\rm inv}, \\
\label{eq:O16}
d\YY{16}{O} /dt &= \YY{13}{C} Y(\alpha) \rr{5}, \\
 \label{eq:Ne20}
d\YY{20}{Ne} /dt &= \frac{\YYSQ{12}{C}}{2} \rr{2}, \\
 \label{eq:Na23}
d\YY{23}{Na} /dt &= \frac{\YYSQ{12}{C}}{2} \rr{1} - \YY{23}{Na} \lambda_7 +
\YY{23}{Ne} \lambda_8, \\
\label{eq:Ne23}
d\YY{23}{Ne} /dt &= \YY{23}{Na} \lambda_7 -
\YY{23}{Ne} \lambda_8, \\
\label{eq:protons}
dY(p) /dt &= \frac{\YYSQ{12}{C}}{2} \rr{1} - \YY{12}{C} Y(p) \rr{3} + \YY{13}{N} \lambda_3^{\rm inv}, \\
\label{eq:alphas}
dY(\alpha) /dt &= \frac{\YYSQ{12}{C}}{2} \rr{2} - \YY{13}{C} Y(\alpha)
\rr{5}, \\
\label{eq:neutrons}
dY(n) /dt &= -\YY{12}{C} Y(n) \rr{6} + \YY{13}{C} Y(\alpha) \rr{5},
\end{align}
where $\lambda_3^{\rm inv}$ is the thermally--averaged cross--section of the inverse reaction
\reac{13}{N}{\gamma}{p}{12}{C}.

Numerical experiments with a detailed network show that the mole
fractions of protons, $\alpha$-particles, neutrons and \xx{13}{N}
nuclei will be many orders of magnitude smaller than those of
\xx{12}{C}, \xx{16}{O}, \xx{20}{Ne} and, depending on the density and
temperature, \xx{13}{C} and \xx{23}{Na} or \xx{23}{Ne}. This means
that their evolution is very fast compared to that of the rest of the
species, even when very small quantities of \xx{12}{C} have been
burnt. Hence, we assume that the l.h.s. of equations~(\ref{eq:13N}),
(\ref{eq:protons}), (\ref{eq:alphas}) and (\ref{eq:neutrons}) is much
smaller than each individual term in the r.h.s. of their
respective equations. Neglecting these time derivatives we can write
the following equilibrium mole fractions for the nuclei
$p,\ \alpha,\ n$ and \xx{13}{N}, which will hereafter be referred to as
the \emph{trace nuclei}:
\begin{align}
\bar Y(p) &= \YY{12}{C}~ \frac{\lambda_1}{2 \lambda_3} f_{\rm inv}, &  \bar Y(\alpha) &= \frac{\YYSQ{12}{C}}{\YY{13}{C} }~ \frac{\lambda_2}{2 \lambda_5}, \notag \\
\bar Y(n) &= \YY{12}{C}~ \frac{\lambda_2}{2 \lambda_6}, & \YYb{13}{N} &= \YYSQ{12}{C}~ \frac{\rr{1}}{2 \lambda_4}, \label{eq:approx_proton}
\end{align}
where $f_{\rm inv}$, defined as $f_{\rm inv} \equiv 1 + \lambda_3^{\rm
  inv}/\lambda_4$, indicates the relative strength of the inverse
reaction \reac{13}{N}{\gamma}{p}{12}{C} with respect to \xx{13}{N}
e$^-$--captures. This factor is independent of the composition and we
will normally have $f_{\rm inv} \approx 1$, except if we approach
ignition at relatively low densities, where \xx{13}{N} e$^-$--captures
cannot compete with the inverse reaction \reac{13}{N}{\gamma}{p}{12}{C},
e.g. in off--center ignition.

If small quantities of \xx{12}{C} are burnt, the mole fractions of the
trace nuclei will reach the equilibrium values in
equations~(\ref{eq:approx_proton}). The typical time--scales for the
equilibrium values to be reached from either lower or higher
abundances can be found dividing equations~(\ref{eq:approx_proton}) by
the positive terms in the r.h.s. of equations~(\ref{eq:protons}),
(\ref{eq:alphas}), (\ref{eq:neutrons}) and (\ref{eq:13N}), or an
arbitrary higher--than--equilibrium mole fraction by the negative terms
in the latter equations. Both calculations give the same time--scales,
assuming $f_{\rm inv} = 1$, namely:
\begin{align}
\tau(p) = \bigl[\YY{12}{C} \rr{3}\bigr]^{-1}, & \ \ \ \ \ \ \ \tau(\alpha) = \bigl[\YY{13}{C} \rr{5}\bigr]^{-1} \notag \\
\tau(n) = \bigl[\YY{12}{C} \rr{6}\bigr]^{-1}, & \ \ \ \ \ \ \ \tau(\xxf{13}{N}) = \lambda_4^{-1}. \label{eq:approx_tau}
\end{align}
In contrast, the typical time--scale for \xx{12}{C} burning will be
\begin{align}
  \tau(\xxf{12}{C}) \approx \bigl[\YY{12}{C} \rho N_{\rm A} (\lambda_1 + \lambda_2)\bigr]^{-1},
\end{align}
and the typical time--scales for \xx{23}{Na} electron--captures or
\xx{23}{Ne} $\beta$--decays will be $\lambda_7^{-1}$ and
$\lambda_8^{-1}$, respectively. We compute these time--scales at the
temperatures and densities relevant for white dwarf interiors (see
Table~\ref{tab:timescales}) and find that the following relations
will be normally satisfied:
\begin{align} \label{eq:ineq}
  \tau(n) < \tau(p) < \tau(\alpha)  <  \tau(\xxf{13}{N})  < \tau(\xxf{12}{C})
\end{align}

Given that the convective time--scales found in pre--ignition white
dwarfs will be normally bigger than the biggest trace nuclei
time--scale shown in Table~\ref{tab:timescales}, it can be assumed
that the trace nuclei will be in equilibrium even within moving
convective eddies. The trace nuclei equilibrium mole fractions will be
analogous to other equilibrium state variables like the temperature,
which is defined by the assumption of local thermodynamic equilibrium
inside every point of the star. In fact, in every stellar evolution
code, as far as the authors are aware, energy diffusion during
convection is computed assuming that local thermodynamic equilibrium
is achieved in time--scales much shorter than the convective
turn--over time--scale.

Thus, we can assume that the trace nuclei equilibrium mole fractions
are given by equations~(\ref{eq:approx_proton}) in every point of the
star. Strictly speaking, the reactions needed to reach the equilibrium
values will break the assumption of energy conservation under this
approximation. However, since the trace nuclei equilibrium mole
fractions will be very small compared to the \xx{12}{C} mole fraction
changes, this effect will be negligible during \xx{12}{C} burning.

\begin{table}\centering
\caption{Characteristic time--scales for a typical CO WD interior
  composition when the \xx{12}{C} burning time--scale is less than
  $10^{10}$ yr. We assumed the lowest \xx{13}{C} abundances found in
  simulations that start with solar abundance in order to compute
  $\tau_\alpha$ . In all cases the trace nuclei characteristic
  time--scales are much smaller than $\tau(\rm ^{12}C)$,
  $\lambda_7^{-1}$, $\lambda_8^{-1}$, or the convective turn--over
  time--scale [$50$ sec before ignition at a density of $1.8\E{9}$ g
    cm$^{-3}$ in the worst case according to
    \citet{2006MNRAS.368..187L}], which validates the use of the
  approximated network in both radiative and convective energy
  transport regions.}
\begin{tabular}{c c | c c c c c c c}
  \\
  \hline
  $\rho$ & T  & $\tau(n)$ & $\tau(p)$ & $\tau(\alpha)$ & $\tau(\rm ^{13}N)$ & $\tau(\rm ^{12}C)$ & $\lambda_7^{-1}$ & $\lambda_8^{-1}$ \\
  {[g cm$^{-3}$]} & [K] & \multicolumn{7}{c}{[s]}\\
  \hline
  1\EE{7} & 5\EE{8} & 1\EE{-10} & 1\EE{-7}  & 7\EE{0}   & 6\EE{1} & 2\EE{13} & 1\EE{47} & 6\EE{1} \\    
          & 7\EE{8} & 1\EE{-10} & 2\EE{-8}  & 4\EE{-2}  & 6\EE{1} & 5\EE{8} & 9\EE{32} & 6\EE{1} \\     
          & 8\EE{8} & 1\EE{-10} & 1\EE{-8}  & 2\EE{-1}  & 6\EE{1} & 1\EE{7} & 9\EE{30} & 6\EE{1} \\     
  \hline                                                                                                %
  1\EE{8} & 5\EE{8} & 1\EE{-11} & 1\EE{-8}  & 3\EE{-1}  & 6\EE{0} & 2\EE{11} & 2\EE{36} & 1\EE{2} \\    
          & 7\EE{8} & 1\EE{-11} & 1\EE{-9}  & 2\EE{-3}  & 6\EE{0} & 1\EE{7} & 5\EE{25} & 1\EE{2} \\     
          & 8\EE{8} & 1\EE{-11} & 7\EE{-10} & 1\EE{-3}  & 6\EE{0} & 3\EE{5} & 3\EE{23} & 1\EE{2} \\     
  \hline                                                                                                %
  1\EE{9} & 5\EE{8} & 1\EE{-12} & 4\EE{-10} & 4\EE{-3}  & 2\EE{-1} & 2\EE{8} & 5\EE{12} & 2\EE{7} \\    
          & 7\EE{8} & 1\EE{-12} & 6\EE{-11} & 4\EE{-5}  & 2\EE{-1} & 4\EE{4} & 7\EE{9} & 2\EE{6} \\     
          & 8\EE{8} & 1\EE{-12} & 4\EE{-11} & 1\EE{-5} & 2\EE{-1} & 2\EE{3} & 2\EE{9} & 8\EE{5} \\     
  \hline                                                                                                %
  3\EE{9} & 3\EE{8} & 4\EE{-13} & 1\EE{-9}  & 1\EE{-2}  & 5\EE{-2} & 6\EE{10} & 3\EE{3} & 3\EE{26} \\   
          & 5\EE{8} & 4\EE{-13} & 5\EE{-11} & 4\EE{-5}  & 5\EE{-2} & 1\EE{6} & 2\EE{3} & 3\EE{20} \\    
          & 7\EE{8} & 4\EE{-13} & 1\EE{-11} & 2\EE{-6} & 5\EE{-2} & 9\EE{2} & 2\EE{3} & 2\EE{15} \\    
          & 8\EE{8} & 4\EE{-13} & 8\EE{-12} & 3\EE{-6} & 5\EE{-2} & 5\EE{1} & 2\EE{3} & 1\EE{14} \\    
  \hline                                                                                                %
  6\EE{9} & 1\EE{8} & 2\EE{-13} & 2\EE{-11} & 2\EE{-3}  & 2\EE{-2} & 1\EE{9} & 8\EE{1} & 7\EE{59} \\    
          & 3\EE{8} & 2\EE{-13} & 1\EE{-10} & 5\EE{-4}  & 2\EE{-2} & 1\EE{8} & 8\EE{1} & 2\EE{50} \\    
          & 5\EE{8} & 2\EE{-13} & 1\EE{-11} & 4\EE{-6} & 2\EE{-2} & 2\EE{4} & 7\EE{1} & 2\EE{39} \\    
          & 7\EE{8} & 2\EE{-13} & 3\EE{-12} & 3\EE{-7} & 2\EE{-2} & 4\EE{1} & 6\EE{1} & 3\EE{27} \\    
          & 8\EE{8} & 2\EE{-13} & 2\EE{-12} & 2\EE{-7} & 2\EE{-2} & 3\EE{0} & 6\EE{1} & 1\EE{25} \\    
  \hline
\end{tabular}
\label{tab:timescales}
\end{table}

It is worth noticing that the trace nuclei equilibrium mole fractions
will change with the $\tau(\xxf{12}{C})$ time--scale. For
example, when \xx{13}{C} nuclei are synthesized by the burning of
\xx{12}{C}, the equilibrium abundance of $\alpha$--particles will
decrease significantly as can be seen in
equations~(\ref{eq:approx_proton}). Assuming that the mole fraction of
\xx{12}{C} is constant for this purpose, the $\alpha$--particle mole
fraction would decrease as $d \ln Y(\alpha) = -d \ln Y(\xxf{13}{C})$.

With the assumption of trace nuclei equilibrium we can obtain a
simplified set of equations which do not include terms with the trace
nuclei typical time--scales. Replacing
equations~(\ref{eq:approx_proton}) into equations~(\ref{eq:C12}),
(\ref{eq:C13}), (\ref{eq:O16}), (\ref{eq:Ne20}), (\ref{eq:Na23}) and
(\ref{eq:Ne23}), we obtain:
\begin{align}
  \label{eq:C12simple}
  d\YY{12}{C}/dt &= -3~ \frac{\YYSQ{12}{C}}{2} \rho N_{\rm A} (\lambda_1 +  \lambda_2), \\
  \label{eq:C13simple}
  d\YY{13}{C}/dt &= \frac{\YYSQ{12}{C}}{2} \rho N_{\rm A} \lambda_1  \\
  \label{eq:O16simple}
  d\YY{16}{O}/dt &= \frac{\YYSQ{12}{C}}{2} \rr{2}, \\
  \label{eq:Ne20simple}
  d\YY{20}{Ne}/dt &= \frac{\YYSQ{12}{C}}{2} \rr{2}, \\
  \label{eq:Na23simple}
  d\YY{23}{Na}/dt &= \frac{\YYSQ{12}{C}}{2} \rr{1} - \YY{23}{Na} \lambda_7 + \YY{23}{Ne}\lambda_8, \\
  \label{eq:Ne23simple}
  d\YY{23}{Ne}/dt &= \YY{23}{Na} \lambda_7 - \YY{23}{Ne}\lambda_8,
\end{align}
which constitutes the system of equations for a simplified nuclear
network.

A quick inspection of these equations shows that only \xx{12}{C},
\xx{23}{Na} and \xx{23}{Ne} need to be tracked as primary species to
accurately follow the chemistry changes in the star under this
approximation, which will greatly simplify the computational cost of
simultaneously solving the structure and chemistry of the
star. Moreover, it can be seen that \xx{12}{C} burns 50\% faster than
what one would naively obtain using only the two main carbon burning
reactions and ignoring the presence of small quantities of ashes, as
first noticed in \citet{2008ApJ...673.1009P} and
\citet{2008ApJ...677..160C}.

We can also see that the evolution of the \xx{13}{C}, \xx{16}{O} and
\xx{20}{Ne} nuclei will be initially faster than that of \xx{12}{C},
but as significant amounts of \xx{12}{C} are burnt their
characteristic time--scales will become comparable, i.e. assuming
$\lambda_1 \approx \lambda_2$:
\begin{align}
&\tau(\xxf{12}{C}) = \frac{2}{3}~ \bigl[\YY{12}{C} \rho N_{\rm A} (\lambda_1 + \lambda_2)\bigr]^{-1} \notag \\
&\tau(\xxf{13}{C}) = 2~\frac{\YY{13}{C}}{\YY{12}{C}}~\bigl[\YY{12}{C} \rr{1}\bigr]^{-1} 
  \approx 6~\frac{\YY{13}{C}}{\YY{12}{C}}~\tau(\xxf{12}{C}), \notag \\ 
&\tau(\xxf{16}{O}) = 2~\frac{\YY{16}{O}}{\YY{12}{C}}~\bigl[\YY{12}{C} \rr{2}\bigr]^{-1}
  \approx 6~\frac{\YY{16}{O}}{\YY{12}{C}}~\tau(\xxf{12}{C}), \notag \\
&\tau(\xxf{20}{Ne}) = 2~\frac{\YY{20}{Ne}}{\YY{12}{C}}~\bigl[\YY{12}{C} \rr{2}\bigr]^{-1}
  \approx 6~\frac{\YY{20}{Ne}}{\YY{12}{C}}~\tau(\xxf{12}{C}), \notag 
\end{align}
The evolution of \xx{23}{Na} and \xx{23}{Ne} will be different and
will depend on the density. Their characteristic time--scales will be:
\begin{align}
\tau(\xxf{23}{Na}) &= \min \biggl\lbrace 2~\frac{\YY{23}{Na}}{\YY{12}{C}}~\bigl[\YY{12}{C} \rr{1}\bigr]^{-1},
~~\lambda_7^{-1}, ~~\frac{\YY{23}{Na}}{\YY{23}{Ne}}~ \lambda_8^{-1} \biggr\rbrace \label{eq:tauNa23}\\
\tau(\xxf{23}{Ne}) &= \min \biggl\lbrace \frac{\YY{23}{Ne}}{\YY{23}{Na}}~ \lambda_7^{-1}, ~~\lambda_8^{-1} \biggr\rbrace \label{eq:tauNa23}
\end{align}

If the \xx{23}{Ne} $\beta$--decay time--scale were much shorter than
the \xx{23}{Na} $e^-$--capture time--scale ($\lambda_8^{-1} \ll
\lambda_7^{-1}$) and the \xx{12}{C}--burning time--scale
       [$\lambda_8^{-1} \ll \tau(\xxf{12}{C})$], which corresponds to
       the low density limit and which we call hypothesis H1,
       \xx{23}{Ne} nuclei produced by $e^-$--captures from newly
       synthesized \xx{23}{Na} would move to an equilibrium value in a
       time--scale $\lambda_8^{-1}$. This value would be obtained
       neglecting time derivatives in equation~(\ref{eq:Ne23simple}):
\begin{align}
  \YYb{23}{Ne} = \YY{23}{Na} \frac{\lambda_7}{\lambda_8},
\end{align}
and using this value in equation~(\ref{eq:Na23simple}) we obtain:
\begin{align}
  d\YY{23}{Na}/dt &= \frac{\YYSQ{12}{C}}{2} \rr{1}. \label{eq:Na23simple2}
\end{align}
If the $e^-$--capture time--scale were much smaller than the
\xx{23}{Ne} $\beta$--decay time--scale ($\lambda_7^{-1} \ll
\lambda_8^{-1}$) and the \xx{12}{C}--burning time--scale
       [$\lambda_7^{-1} \ll \tau(\xxf{12}{C})$], which corresponds to
       the high density limit and which we call hypothesis H2,
       \xx{23}{Na} would act as trace nuclei and with a time--scale
       $\lambda_7^{-1}$ would reach an equilibrium value, which could
       be obtained neglecting time--derivatives in
       equation~(\ref{eq:Na23simple}):
\begin{align}
  \YYb{23}{Na} = \frac{\YYSQ{12}{C}}{2}~ \frac{\rr{1}}{\lambda_7}+ \YY{23}{Ne}~ \frac{\lambda_8}{\lambda_7}.
\end{align}
Assuming this equilibrium value in equation~(\ref{eq:Ne23simple}) we
obtain:
\begin{align}
  d\YY{23}{Ne}/dt &= \frac{\YYSQ{12}{C}}{2} \rr{1} \label{eq:Ne23simple2}
\end{align}

These alternative additional simplifications are useful for
understanding the behavior of the simplified network in these extreme
cases, but are not used in the simplification of this section. They
are not valid if the \xx{12}{C}--burning time--scale becomes smaller
than both the \xx{23}{Na} $e^-$--capture and \xx{23}{Ne}
$\beta$--decay time--scales, or if the latter time--scales are similar
to each other. If this is the case, we must solve
equations~(\ref{eq:Na23simple}) and (\ref{eq:Ne23simple}) exactly
under the simplified network.

\subsection{Dependence on the \xx{12}{C} Burnt Mole Fraction}
We can also derive a set of differential equations that relate the
increase in mole fraction of the secondary species with the amount of
burnt carbon. First, we divide equations~(\ref{eq:C13simple}),
(\ref{eq:O16simple}) and (\ref{eq:Ne20simple}) by
equation~(\ref{eq:C12simple}) to obtain:
\begin{align}
  \label{eq:dC13C12}
  \frac{d\YY{13}{C}}{d\YY{12}{C}} &= -\frac{1}{3(1 + \lambda_2/\lambda_1)} \approx -0.15, \\
  \label{eq:dO16C12}
  \frac{d\YY{16}{O}}{d\YY{12}{C}} &= -\frac{1}{3(1 + \lambda_1/\lambda_2)} \approx -0.19, \\
  \label{eq:dNe20C12}
  \frac{d\YY{20}{Ne}}{d\YY{12}{C}} &= -\frac{1}{3(1 + \lambda_1/\lambda_2)} \approx -0.19.
\end{align}

For \xx{23}{Na} or \xx{23}{Ne} the result is not as simple, unless
either of the conditions necessary for
equations~(\ref{eq:Na23simple2}) or (\ref{eq:Ne23simple2}) are met. In
those cases, we would get
\begin{itemize}
  \item Under hypothesis H1 (low density limit):
    \begin{align}
      \frac{d\YY{23}{Na}}{d\YY{12}{C}} \approx -\frac{1}{3(1 + \lambda_2/\lambda_1)} \approx -0.15 \label{eq:dNa23C12}
    \end{align}
  \item Under hypothesis H2 (high density limit):
    \begin{align}
      \frac{d\YY{23}{Ne}}{d\YY{12}{C}} \approx -\frac{1}{3(1 + \lambda_2/\lambda_1)} \approx -0.15
    \end{align}
\end{itemize}

\subsection{Electron Mole Fraction}

Similarly, we can write equations for the evolution of the electron
mole fraction $Y_{\rm e}$. Note that this quantity is closely related
to the neutron excess, defined as $\eta \equiv \sum X_{\rm i}
\eta_{\rm i}$, with $\eta_{\rm i} \equiv (n_{\rm i} - p_{\rm
  i})/(n_{\rm i} + p_{\rm i})$ and where $n_{\rm i}$ and $p_{\rm i}$
are the number of neutrons and protons of the respective species. The
electron mole fraction and neutron excess are related by the formula
$\eta = 1 - 2 ~Y_{\rm e}$, which implies $d\eta/dt = -2 ~dY_{\rm
  e}/dt$.

We note that when the \xx{12}{C}--burning time--scale is much smaller
than the $e^-$--capture time--scale [$\tau(\xxf{12}{C}) \ll
  \lambda_7^{-1}$] and when \xx{13}{N} is in equilibrium, the rate at
which $Y_{\rm e}$ changes will be the rate at which $\YY{12}{C}$ changes
due to the reaction \reac{12}{C}{^{12}{\rm C}}{p}{23}{Na},
i.e.:
\begin{align} \label{eq:dYeC12}
  \frac{dY_{\rm e}}{dt} = -\frac{\YYSQ{12}{C}}{2} \rr{1}   ~~{\rm and}
  ~~~  \frac{dY_{\rm e}}{d\YY{12}{C}} = \frac{1}{3(1 + \lambda_2 /
    \lambda_1)} \approx 0.15~.
\end{align}
When the $e^-$--capture time--scale becomes smaller than the
\xx{12}{C} burning time--scale, the $e^-$--capture rate will be
twice the former result, since \xx{23}{Na} is produced in the same
branch of the network as \xx{13}{N} (see
Fig.~\ref{fig:treeC12high}). If both time--scales are comparable, the
electron mole fraction can be obtained using that $Y_{\rm e} = (1 -
\eta) / 2$ and computing the neutron excess from the exact solution of
equations~(\ref{eq:C12simple}) to (\ref{eq:Ne23simple}) and the formula
\begin{align}
  \eta \approx Y(\xxf{13}{C}) + Y(\xxf{23}{Na}) + 3~ Y(\xxf{23}{Ne})  
  + \bar Y(n) - \bar Y(p) - \bar Y(\xxf{13}{N}).
\end{align}

\section{LIMITS OF THE SIMPLIFIED NETWORK N1}
Now we will examine different ways the assumptions of the simplified
network N1 can break down, and with this information build a second
more refined simplified network, N2, which includes some of the leak
reactions that will be discussed in what follows.

\subsection{Proton Leaks} \label{sec:protonleak}

\begin{figure}
  \centering
  \includegraphics[width=0.7\hsize,angle=270,keepaspectratio]{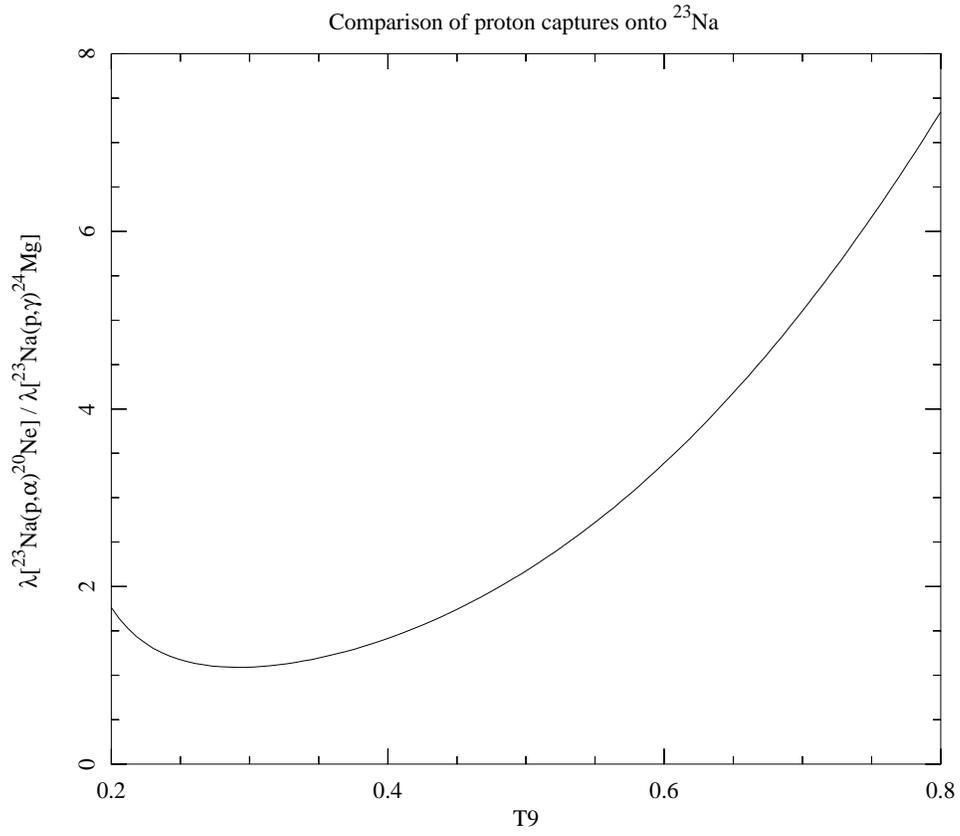}
  \caption[Proton capture to \xx{23}{Na}]{Cross--section ratio for
    proton captures onto \xx{23}{Na}. The dominant proton--leak
    reaction will be \reac{23}{Na}{p}{\alpha}{20}{Ne} for all the
    temperature range considered here.}
  \label{fig:comparison_23Naproton}
\end{figure}

When the \xx{23}{Ne} $\beta$--decay time--scale is shorter than the
\xx{23}{Na} $e^-$--capture time--scale, normally below the threshold
density $\rho_{\rm th}\approx 1.7$ g cm$^{-3}$, some of the protons
that would be captured in the \reac{12}{C}{p}{\gamma}{13}{N} reaction
can leak via the reactions \reac{23}{Na}{p}{\alpha}{20}{Ne} and
\reac{23}{Na}{p}{\gamma}{24}{Mg}, which are defined as reactions 9) and
10), as the abundance of \xx{23}{Na} increases.  The ratio of their
thermally averaged cross--sections is shown in
Figure~\ref{fig:comparison_23Naproton}.

\begin{figure}
  \centering
  \includegraphics[width=0.7\hsize,angle=270,keepaspectratio]{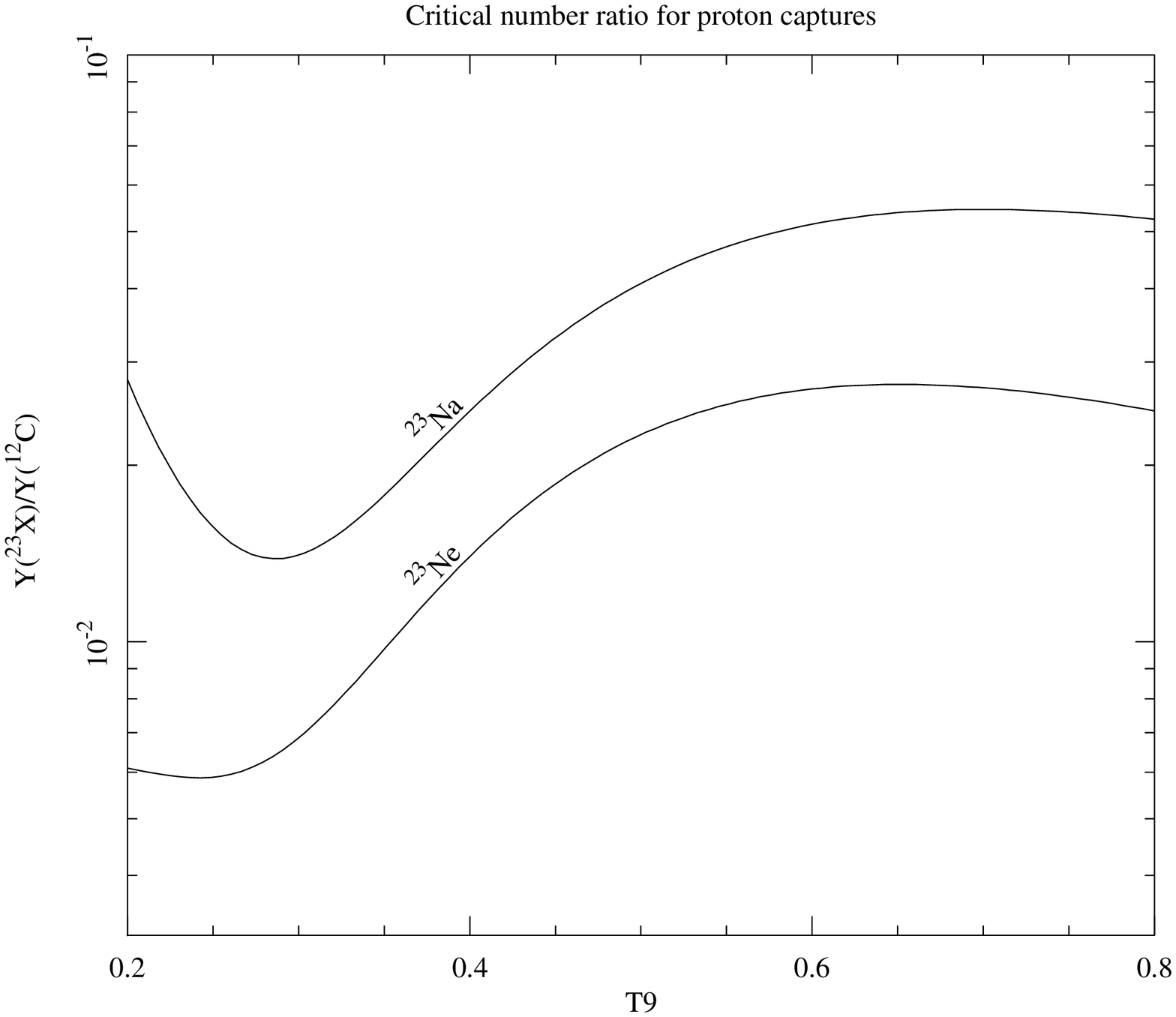}
  \caption[Proton capture critical abundances]{Mole fraction ratio
    between Urca matter and \xx{12}{C} when proton captures onto Urca
    matter become equal to proton captures onto \xx{12}{C}, plotted
    against the temperature in units of $10^9$ K. Screening
    corrections are taken into account, using a density of 1$\E{9}$ g
    cm$^{-3}$ for \xx{23}{Na} and 2$\E{9}$ g cm$^{-3}$ for
    \xx{23}{Ne}, below and above the threshold for electron captures.}
  \label{fig:pcapX23}
\end{figure}

Conversely, when the \xx{23}{Na} $e^-$--capture time--scale is shorter
than the \xx{23}{Ne} $\beta$--decay time--scale, normally above the
threshold density $\rho_{\rm th}\approx 1.7$ g cm$^{-3}$, protons can
leak via the reaction \reac{23}{Ne}{p}{n}{23}{Na}, which is defined as
reaction 11), as the abundance of \xx{23}{Ne} increases. These leak
reactions can significantly change the distribution of ashes and
energy input as \xx{12}{C} is burnt, as well as the trace nuclei
equilibrium mole fractions, as will be shown later. The abundances of
\xx{23}{Na} and \xx{23}{Ne} at which the former reaction rates become
equal to the \reac{12}{C}{p}{\gamma}{13}{N} reaction are shown in
Figure~\ref{fig:pcapX23}.

Let us define the cross--section for proton captures on either
nuclei of the pair \xx{23}{Na}--\xx{23}{Ne}, which we call Urca
matter, as $\lambda_{\rm U}$, and the cross--section for proton
captures on \xx{12}{C} as $\lambda_{\rm C}$. The ratio between the
rates in both reactions will be:
\begin{equation}
  r = \frac{Y(\xxf{23}{Na}~~\rm{or}~~\xxf{23}{Ne})}{Y(\xxf{12}{C})}~\frac{\lambda_{\rm U}}{\lambda_{\rm C}}.
\end{equation}
We can use the following approximate relation inferred from the
simplified version of the network:
\begin{align}
  \frac{dY(\xxf{23}{Na}~~\rm{or}~~\xxf{23}{Ne})}{dY(\xxf{12}{C})} &\approx -0.15
\end{align}
and defining $f$ as the burnt fraction of \xx{12}{C}, we can write:
\begin{align}
  r \approx \frac{-0.15 ~ \Delta
    Y(\xxf{12}{C})}{Y(\xxf{12}{C})}~\frac{\lambda_{\rm
      U}}{\lambda_{\rm C}} = 0.15~ f ~\frac{\lambda_{\rm
      U}}{\lambda_{\rm C}},
\end{align}
i.e. when $r=1$, $f \approx 6.7 ~\lambda_{\rm C} / \lambda_{\rm U}$,
or $\approx 6.7 \times$ the ratio shown in
Figure~\ref{fig:pcapX23}. Thus, when 9\% or 4\% of the original
\xx{12}{C} is burnt, depending on whether protons leak on \xx{23}{Na}
or \xx{23}{Ne}, proton leaks will become significant. From
Figure~\ref{fig:pcapX23} it can be noted that proton leaks will be
stronger at temperatures of about $3\E{8}$ K. Above temperatures of
$5\E{8}$ K the fraction of burnt \xx{12}{C} for proton leaks to be
important will change to approximately 27\% and 14\%, when either the
\xx{23}{Ne} $\beta$--decay time--scale or the \xx{23}{Na} electron
capture time--scale is shorter, respectively.

Although proton leaks start later when the \xx{23}{Ne} $\beta$--decay
time--scale is shorter, their effect in this case is more difficult to
model. Since less \xx{13}{N} will be present due to the bypassing of
one of the branches of the network, the amount of $e^-$--captures is
reduced and less $\alpha$--capturing \xx{13}{C} synthesised, opening
secondary channels for $\alpha$--captures. With more proton leaks,
\xx{12}{C} and \xx{16}{O} can become the main targets for
$\alpha$--captures, depending on the temperature as will be discussed
later.

When the \xx{23}{Na} $e^-$--capture time--scale is shorter, proton
leaks will start sooner, which does not change significantly the main
results of the simplified network because the end products will be the
same after two $e^-$--captures and the burning of six \xx{12}{C}
nuclei:
\begin{align}
  6~\nn{12}{C} + 2~e^- ~\rightarrow~ \nn{23}{Ne}~ +~ \nn{20}{Ne}~ +~
  \nn{16}{O}~ +~ \nn{13}{C}.
\end{align}
The main difference will be that now both $e^-$--captures will be
on \xx{23}{Na} rather than on \xx{13}{N} and \xx{23}{Na}, changing
the typical time--scale of the network in this leak branch.

\begin{figure}[ht!]
  \centering
  \includegraphics[width=0.8\hsize,angle=270,keepaspectratio]{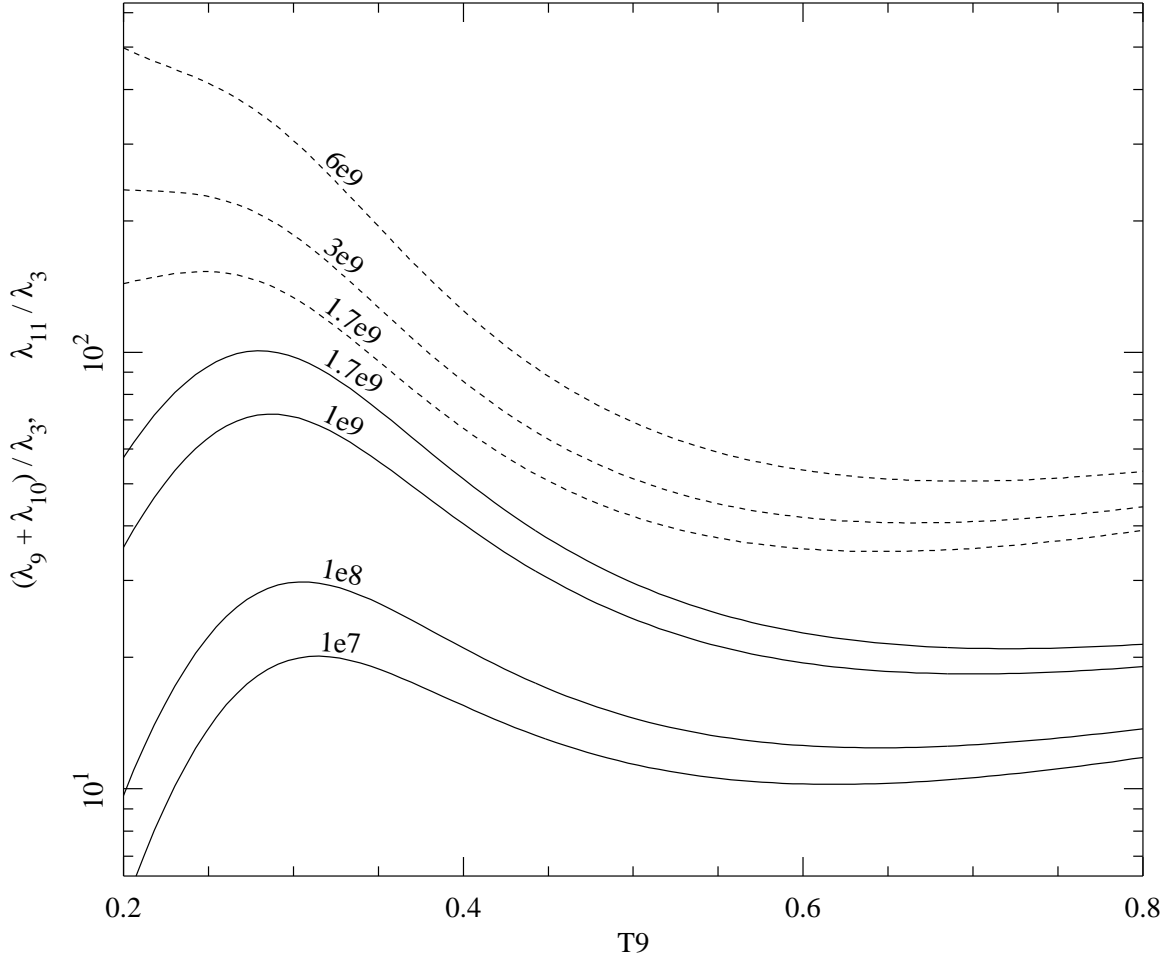}
  \caption[Main \xx{12}{C} burning reactions]{Ratio between thermally
    averaged cross--sections relevant for proton captures at different
    densities [gr~cm$^{-3}$]. Solid lines correspond to $(\lambda_9 +
    \lambda_{10}) / \lambda_3$ and dashed lines, to $\lambda_{13} /
    \lambda_3$ (see Section~\ref{sec:protonleak}). The threshold density for $e^-$--captures onto
    \xx{23}{Na} is close to 1.7$\E{9}$ g cm$^{-3}$. The density
    dependence is due to screening corrections.}
  \label{fig:comparison_proton2}
\end{figure}

\begin{figure*}[ht!]
  \centering
  \includegraphics[width=0.7\hsize,keepaspectratio]{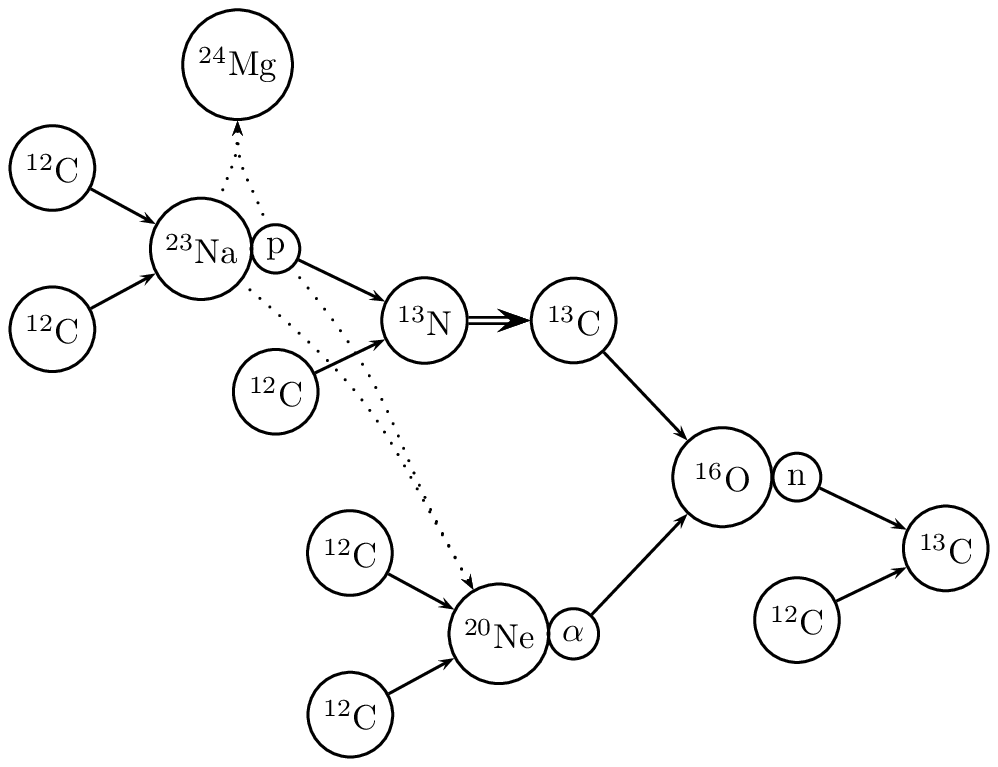}
         \caption[Schematic low density \xx{12}{C} burning with proton
           leaks]{Same as Figure~\ref{fig:treeC12low}, but with
           proton leaks (dotted arrows). See discussion in the text
           for neutron and $\alpha$--leaks.}
         \label{fig:treeC12low+leak}
\end{figure*}

\begin{figure*}[ht!]
  \centering
  \includegraphics[width=0.7\hsize,keepaspectratio]{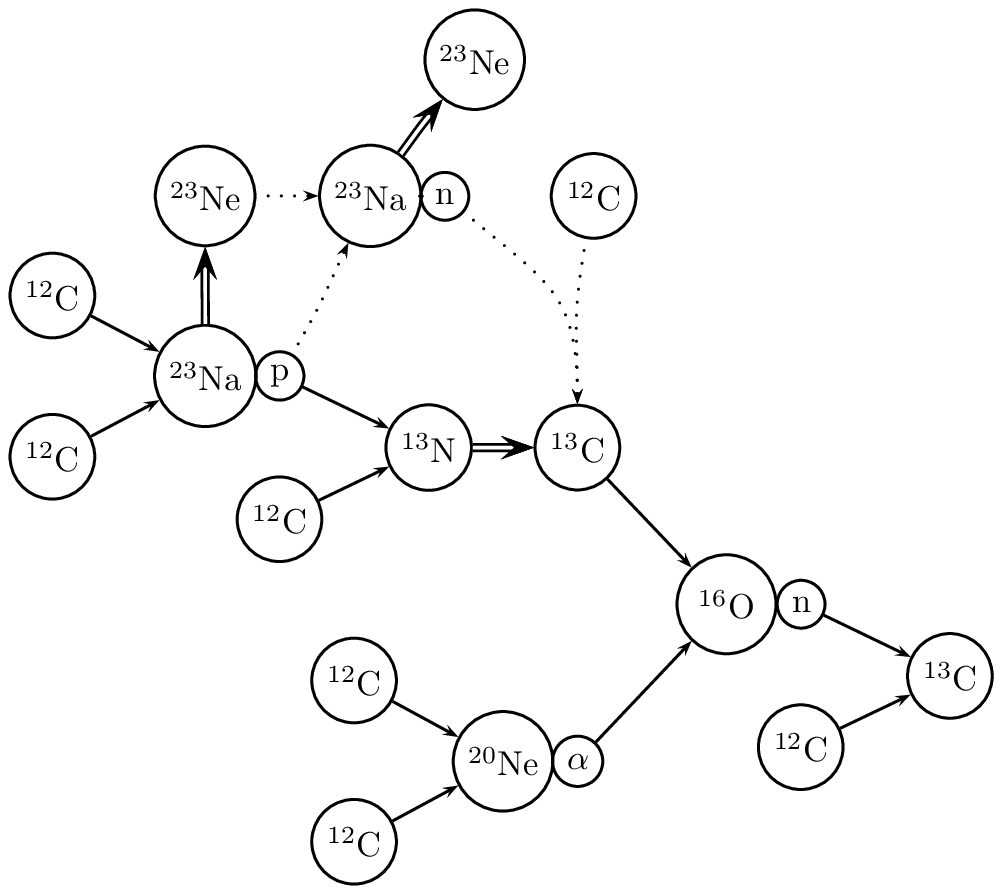}
         \caption[Schematic high density \xx{12}{C} burning with
           proton leaks]{Same as Figure~\ref{fig:treeC12high}, but
           with proton leaks and the resulting bypass for
           neutron captures (dotted arrows). See discussion in the
           text for neutron and $\alpha$--leaks.}
         \label{fig:treeC12high+leak}
\end{figure*}

The ratios between the relevant thermally averaged cross--sections for
proton captures at different densities are shown in
Figure~\ref{fig:comparison_proton2}. We summarize the proton--leak
reactions in Figures~\ref{fig:treeC12low+leak} and
\ref{fig:treeC12high+leak}.


\subsection{Neutron Leaks} \label{sec:neutronleak}
A similar analysis can be done to compute the critical mole fraction
of secondary nuclei for neutron leaks to be important. In this case,
the most important neutron capture reactions are
\reac{12}{C}{n}{\gamma}{13}{C}, \reac{20}{Ne}{n}{\gamma}{21}{Ne},
\reac{21}{Ne}{n}{\gamma}{22}{Ne} and
\reac{23}{Na}{n}{\gamma}{24}{Na}. The \xx{20}{Ne} and \xx{21}{Ne}
neutron capture reactions have very similar cross--sections,
approximately seven times the \xx{12}{C} neutron capture reaction. The
\xx{23}{Na} reaction has a cross--section approximately 11 times bigger
than the cross--section of the \xx{12}{C} neutron capture reaction.

The ratio of the mole fractions of \xx{20}{Ne} and \xx{23}{Na} with
respect to \xx{12}{C} can be related to the mole fraction change of
\xx{12}{C} using equations~(\ref{eq:dNe20C12}) and
(\ref{eq:dNa23C12}), at densities when the \xx{23}{Ne} $\beta$--decay
time--scale is shorter than the \xx{23}{Na} $e^-$--capture
time--scale:
\begin{align} \label{eq:0.34}
  Y(\xxf{20}{Ne}) + Y(\xxf{23}{Na}) &\approx 
  -\Delta Y(\xxf{12}{C}) ~\biggl\lbrace \frac{dY(\xxf{20}{Ne})}{dY(\xxf{12}{C})}
  + \frac{dY(\xxf{23}{Na})}{dY(\xxf{12}{C})}\biggr\rbrace \notag \\
  &\approx -0.34 ~\Delta Y(\xxf{12}{C}),
\end{align}
or when the \xx{23}{Na} $e^-$--capture is shorter than the
\xx{23}{Ne} $\beta$--decay time--scale:
\begin{align} \label{eq:0.19}
  Y(\xxf{20}{Ne}) + Y(\xxf{23}{Na}) \approx -0.19 ~ \Delta
  Y(\xxf{12}{C}).
\end{align}
In the first case, using equation~(\ref{eq:0.34}) and using the added
cross--sections of \xx{20}{Ne}, \xx{21}{Ne} and \xx{23}{Na} neutron
captures, it can be shown that neutron leaks, i.e. significant neutron
captures on species different than \xx{12}{C}, occur when the fraction
of burnt carbon is approximately 12\% its original amount. In the
second case, using equation~(\ref{eq:0.19}) it can be shown that
neutron leaks will occur when about 38\% of the original carbon is burnt.

As first noticed by \citet{2007ApJ...655L..93C}, neutron captures onto
\xx{56}{Fe} will be negligible, since the
\reac{56}{Fe}{n}{\gamma}{57}{Fe} reaction has a cross--section
approximately 64 times bigger than that of neutron capture onto
\xx{12}{C}, but the mole fraction of \xx{56}{Fe} is approximately 1250
times smaller than that of \xx{12}{C} at solar metallicity in the
\xx{12}{C}--rich environment of a WD. Thus, a \xx{56}{Fe} abundance of
more than 20 times the solar abundance would be required to compete
with the reaction \reac{12}{C}{n}{\gamma}{13}{C}.

\subsection{$\alpha$--particle Leaks} \label{sec:alphaleak}
\begin{figure}
  \centering
  \includegraphics[width=0.7\hsize,angle=270,keepaspectratio]{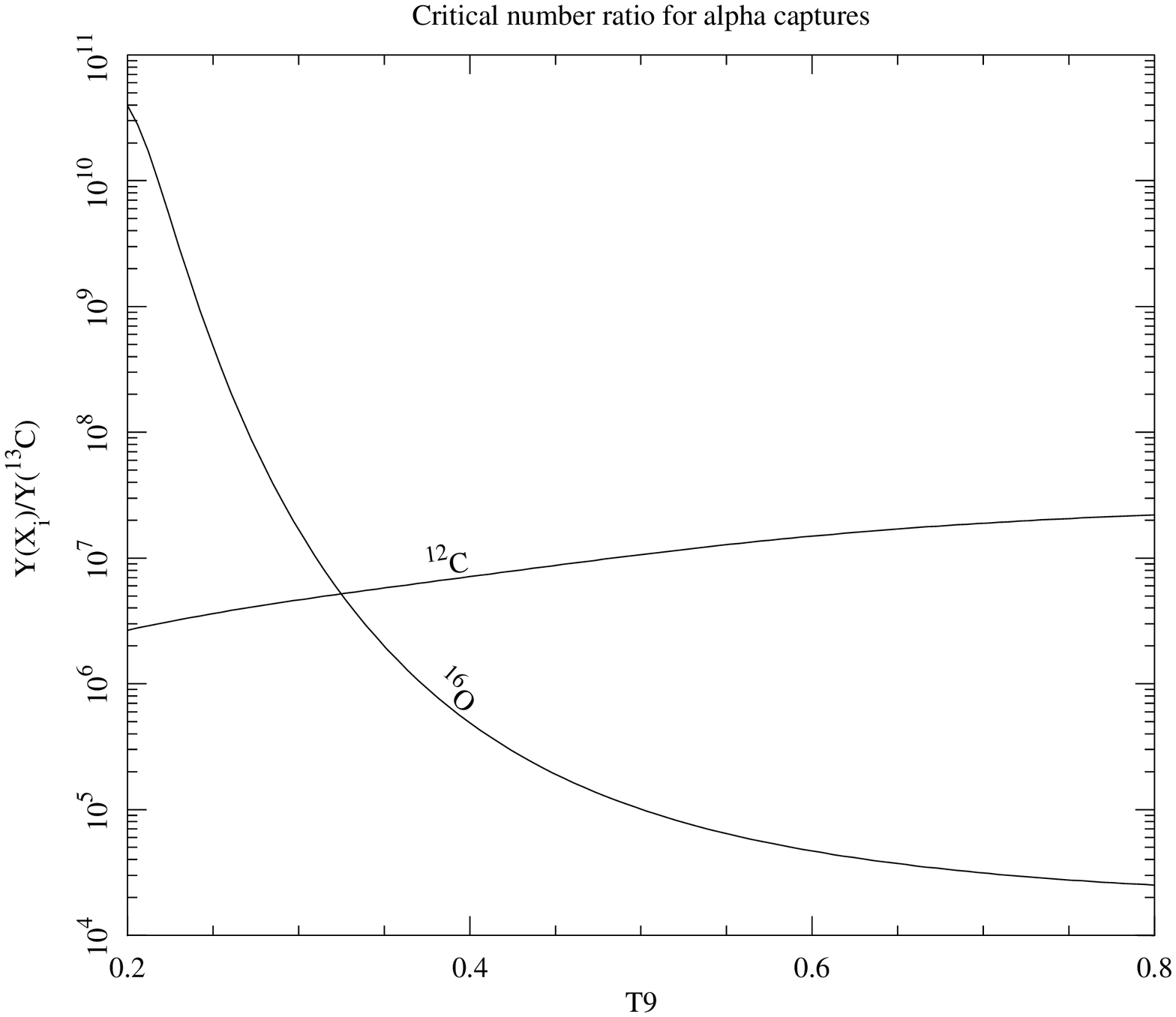}
  \caption[Alpha captures critical abundances]{Number ratio of
    \xx{12}{C} or \xx{16}{O} to \xx{13}{C} when the rates of the
    $\alpha$--capture reactions \reac{12}{C}{\alpha}{\gamma}{16}{O} or
    \reac{16}{O}{\alpha}{\gamma}{20}{Ne} equals the rate of
    $\alpha$--captures in the reaction
    \reac{13}{C}{\alpha}{n}{16}{O}.}
  \label{fig:alphacap}
\end{figure}

Our detailed nuclear network shows that the most important reactions
for $\alpha$--captures are \reac{13}{C}{\alpha}{n}{16}{O}, followed by
the much weaker reactions \reac{12}{C}{\alpha}{\gamma}{16}{O} and
\reac{16}{O}{\alpha}{\gamma}{20}{Ne}. In Figure~\ref{fig:alphacap} we
show the ratios of the mole fractions of \xx{12}{C} and \xx{16}{O}
relative to the mole fraction of \xx{13}{C} necessary for
\reac{13}{C}{\alpha}{n}{16}{O} to be the dominant $\alpha$--capture
reaction. It can be seen that only a minimal amount of \xx{13}{C} is
needed for this to be the case: $Y(\xxf{13}{C}) \approx 3\E{-8}$,
which is $\approx 10^4$ times lower than the solar metallicity value.

Moreover, using arguments similar to those used for proton and neutron
leaks, a fraction of only $4\E{-6}$ of the original \xx{12}{C} needs
to be burnt to reproduce this abundance with zero metallicity,
assuming that for every six \xx{12}{C} nuclei one \xx{13}{C} nucleus
is produced. Thus, under the temperature range investigated here, one
would conclude that these secondary reactions are negligible.

However, if significant neutron leaks occur, \xx{13}{C} can be
significantly depleted and these reactions can become important. In
fact, after \xx{13}{C} is depleted as a consequence of neutron leaks
and the $\alpha$--particle over--production accompanying the
\reac{23}{Na}{p}{\alpha}{20}{Ne} proton leak reaction, $\alpha$--leaks
are necessary to reproduce the abundances of \xx{16}{O}, \xx{20}{Ne}
and to limit the growth of $\alpha$--particles and reproduce the
\xx{13}{C} depletion more accurately.

\clearpage

\section{THE SIMPLIFIED NETWORK: SECOND APPROXIMATION (N2)} \label{sec:2ndapprox}

We can now include the leak reactions discussed in the previous
section and additional ones with the following indices:

\begin{tabular}{@{}l l l l l l @{}}
9)& \reac{23}{Na}{p}{\alpha}{20}{Ne}, & Q = 2.38 MeV, & 10)& \reac{23}{Na}{p}{\gamma}{24}{Mg}, & Q = 11.69 MeV, \\
11)& \reac{23}{Ne}{p}{n}{23}{Na}, & Q = 3.59 MeV, & 12)&  \reac{20}{Ne}{n}{\gamma}{21}{Ne},  & Q = 6.76 MeV, \\
13)& \reac{23}{Na}{n}{\gamma}{24}{Na}, & Q = 6.96 MeV, &14)& \reac{12}{C}{\alpha}{\gamma}{16}{O}, & Q = 7.16 MeV, \\
15)& \reac{16}{O}{\alpha}{\gamma}{20}{Ne}, & Q = 2.84 MeV, &16)& \reac{21}{Ne}{n}{\gamma}{22}{Ne},  & Q = 10.36 MeV, \\
17)& \reac{13}{C}{p}{\gamma}{14}{N}, & Q = 7.55 MeV, &18)& \reac{21}{Ne}{p}{\gamma}{22}{Na},  & Q = 6.74 MeV, \\
19)& \reac{20}{Ne}{\alpha}{\gamma}{24}{Mg}, & Q =  9.32 MeV, &20)& \reac{23}{Ne}{\alpha}{n}{26}{Mg},  & Q =  5.41 MeV, \\
21)& \reac{23}{Na}{\alpha}{p}{26}{Mg}, & Q = 1.82 MeV, &22)& \reac{21}{Ne}{\alpha}{n}{26}{Mg},  & Q = 4.84 MeV, \\
\end{tabular}
\vspace{.5cm}

This list includes proton--leak reactions (9, 10, 11, 17 and 18),
neutron--leak reactions (12, 13 and 16), and $\alpha$--leak reactions
(14, 15, 19, 20, 21 and 22). Proton--leaks are the first to be
significant as \xx{12}{C} ashes are produced, but with more ashes
produced neutron--leaks become relevant too. When this happens,
\xx{13}{C} is rapidly depleted and $\alpha$--leaks become important as
well.

As can be seen from the reactions in the list, it would be necessary
to increase the number of species to account for all proton, neutron
and $\alpha$--leaks exactly. To do this we can either account for
\xx{24}{Mg}, \xx{21}{Ne}, \xx{24}{Na}, \xx{22}{Ne}, \xx{14}{N},
\xx{22}{Na} and \xx{26}{Mg} independently, or we can group some or all
of them into an auxiliary variable. Since reactions 16) and 18) depend
on the abundance of \xx{21}{Ne} nuclei, we solve for \xx{21}{Ne}
independently and group \xx{24}{Mg}, \xx{24}{Na}, \xx{22}{Ne},
\xx{14}{N}, \xx{22}{Na} and \xx{26}{Mg} into a dummy \emph{leak}
species.

Thus, while keeping the number of independent variables small and
ensuring mass conservation, we can introduce $Y_{\rm L}$ as the mole
fraction of a dummy nuclei that represents leaks on \xx{24}{Mg},
\xx{24}{Na}, \xx{22}{Ne}, \xx{14}{N}, \xx{22}{Na} and \xx{26}{Mg}. We
assume that this dummy species has approximately the average mass
number of the species it represents, defining it as:
\begin{align} \label{eq:YL}
  Y_{\rm L} \equiv \max \biggl\lbrace 0,~ \frac{1 - \sum A_{\rm i}
    Y_{\rm i}}{22} \biggr\rbrace
\end{align}
where $A_{\rm i}$ is the mass number of the respective species and the
sum is made over \xx{12}{C}, \xx{13}{C}, \xx{13}{N}, \xx{16}{O},
\xx{20}{Ne}, \xx{21}{Ne}, \xx{23}{Ne} and \xx{23}{Na}. Since the leak
nuclei \xx{24}{Mg}, \xx{24}{Na}, \xx{22}{Ne}, \xx{14}{N} and
\xx{22}{Na} can subsequently capture neutrons with a cross--sections
similar to that of \xx{20}{Ne}, for simplicity we will assume that
leak nuclei capture neutrons with the \xx{20}{Ne} cross--section.

If we now add new terms associated to the $p$, $n$ and $\alpha$--leak
reactions discussed above into equations~(\ref{eq:C12}) to
(\ref{eq:neutrons}) and assume a stationary solution for the trace
nuclei $p,~ \alpha,~ n~\rm{and}~ \xxf{13}{N}$, the modified
equilibrium mole fractions can be described with the following
equations (c.f. equations.~\ref{eq:approx_proton}):
\begin{align}
  \tilde Y(p)  &= \bar Y(p) ~  f_{\rm p}, 
  &\tilde Y(\alpha) &= \bar Y(\alpha) ~  f_{\alpha}, \notag \\
  \tilde Y(n)  &= \bar Y(n) ~  f_\alpha~ f_{\rm n}, 
  &\YYt{13}{N} &= \YYb{13}{N} ~  f_{\rm p}, \label{eq:equilibrium2}
\end{align}
where the auxiliary variable $f_{\rm p}$, $f_{\rm n}$ and $f_\alpha$
are defined as follows:
\begin{align}
  \label{eq:fp}
  f_{\rm p} &\equiv \biggl\lbrace 1 
  + f_{\rm inv} ~\frac{\YY{23}{Na} (\lambda_9 + \lambda_{10}) 
    +\YY{23}{Ne} \lambda_{11} + \YY{13}{C} \lambda_{17} +\YY{21}{Ne}\lambda_{18}
  }{\YY{12}{C} \lambda_3} \biggr\rbrace^{-1} \\
  K &\equiv 1 + \frac{\YY{12}{C} \lambda_{14} + \YY{16}{O} \lambda_{15} + \YY{20}{Ne} \lambda_{19} + \YY{23}{Ne} \lambda_{20} + \YY{23}{Na} \lambda_{21} + \YY{21}{Ne} \lambda_{22}}{\YY{13}{C} \lambda_5} \\
  \label{eq:falpha}
  f_{\alpha} &\equiv K^{-1} ~ \biggl\lbrace 1 
  + 2~ \frac{\YY{23}{Na} \tilde Y(p)}{\YYSQ{12}{C}}~ \frac{\lambda_{9}}{\lambda_2} \biggr\rbrace  \\
  \label{eq:fn}
  f_{\rm n} &\equiv \biggl\lbrace 1 
  + \frac{\YY{23}{Ne} \bigl[\tilde Y(p) \lambda_{11} + \tilde Y(\alpha) \lambda_{20} \bigr] + \YY{21}{Ne} Y(\alpha) \lambda_{22}}{\YY{13}{C} \tilde Y(\alpha) \lambda_5} \biggr\rbrace \notag \\
 &~~~~  \biggl\lbrace 1 + \frac{\bigl[\YY{20}{Ne} + Y_{\rm L}\bigr] \lambda_{12} 
    + \YY{21}{Ne} \lambda_{16} + \YY{23}{Na} \lambda_{13}}{\YY{12}{C} \lambda_6} \biggr\rbrace^{-1},
\end{align}
where we have neglected the contribution to $f_{\rm p}$ from the leak
reaction \reac{23}{Na}{\alpha}{p}{26}{Mg}, which is generally
small. An exact solution that takes into account the additional
protons from this or similar reactions can be obtained by multipliying
the trace abundances $\tilde Y(p)$ and $\YYt{13}{N}$ by the factor
$\lbrace 1 + A \bar Y(\alpha)\rbrace ~\lbrace 1 - A \bar Y(\alpha) B
\tilde Y(p) \rbrace^{-1}$, where $A \equiv 2 \YY{23}{Na} \lambda_{21}
~\lbrace \YYSQ{12}{C} \lambda_1 \rbrace^{-1}$ and $B \equiv 2 K^{-1}
\YY{23}{Na} \lambda_9 ~ \lbrace\YYSQ{12}{C} \lambda_2
\rbrace^{-1}$. The factors $f_\alpha$ and $f_{\rm n}$ should be
computed by using these modified values.

These formulae can be used to approach the leak regime of the network
and can be easily generalized to include more leak reactions involving
protons, neutrons or $\alpha$--particles. They predict a decrease in
the number of free protons and \xx{13}{N} nuclei at equilibrium and
the fact that this decrease starts later when the \xx{23}{Ne}
$\beta$--decay time--scale is shorter than the \xx{23}{Na}
$e^-$--capture time--scale, at low densities. Also, they explain the
contribution to $\alpha$--particles from the proton leak reaction
\reac{23}{Ne}{p}{\alpha}{20}{Ne}, the $\alpha$--leaks from \xx{12}{C}
and \xx{16}{O} captures, the contribution to neutrons from the proton
leak reaction \reac{23}{Ne}{p}{\alpha}{20}{Ne} and the neutron
depletion due to \xx{20}{Ne}, \xx{21}{Ne} and \xx{23}{Na} neutron
captures under our approximation, as well as a simple estimate of the
neutron captures on \xx{24}{Mg}, \xx{24}{Na}, \xx{22}{Ne},
\xx{14}{N} and \xx{22}{Na}.

Now, we can rewrite the differential equations describing the
evolution of the slowly varying nuclei \xx{12}{C}, \xx{13}{C},
\xx{16}{O}, \xx{20}{Ne}, \xx{23}{Na} and \xx{23}{Ne}, using the new
trace nuclei equilibrium values and including the leak reactions
discussed above, i.e.:
\begin{align}
 \label{eq:C12new}
&\frac{d\YY{12}{C}}{dt} = -\YY{12}{C} \rho
N_{\rm A} \biggl\lbrace\YY{12}{C} (\lambda_1 + \lambda_2) + \frac{\tilde Y(p)\lambda_3}{f_{\rm inv}} + \tilde Y(n)
\lambda_6 + \tilde Y(\alpha) \lambda_{14} \biggr\rbrace, \\
\label{eq:C13new}
&\frac{d\YY{13}{C}}{dt} = \YY{13}{N} \lambda_4 + \rho N_{\rm A} \biggl\lbrace \YY{12}{C} \tilde Y(n) \lambda_6
- \YY{13}{C} \bigl[\tilde Y(\alpha) \lambda_5 + \tilde Y(p) \lambda_{17} \bigr] \biggr\rbrace ,\\
\label{eq:O16new}
&\frac{d\YY{16}{O}}{dt} = \rho N_{\rm A} \tilde Y(\alpha) \biggl \lbrace \YY{13}{C}  \lambda_5
+  \YY{12}{C} \lambda_{14} - \YY{16}{O} \lambda_{15} \biggr\rbrace , \\
 \label{eq:Ne20new}
&\frac{d\YY{20}{Ne}}{dt} = \rho N_{\rm A} \biggl\lbrace \frac{\YYSQ{12}{C}}{2} \lambda_2 + \YY{23}{Na} \tilde Y(p) \lambda_{9} - \YY{20}{Ne} \bigl[\tilde Y(n) \lambda_{12}+ \tilde Y(\alpha) \lambda_{19}\bigr] \notag \\
& ~~~~~~~~~~~~~  
+ \YY{16}{O} \tilde Y(\alpha) \lambda_{15} \biggr\rbrace, \\
 \label{eq:Na23new}
&\frac{d\YY{23}{Na}}{dt} = - \YY{23}{Na} \lambda_7 +
\YY{23}{Ne} \lambda_8 + \rho N_{\rm A} \biggl\lbrace \frac{\YYSQ{12}{C}}{2} \lambda_{1}  + \YY{23}{Ne} \tilde Y(p) \lambda_{11}  
\notag \\
&~~~~~- \YY{23}{Na} \bigl[ \tilde Y(p) (\lambda_{9} + \lambda_{10}) + \tilde Y(n) \lambda_{13} + \tilde Y(\alpha) \lambda_{21}\bigr]
\biggr\rbrace, \\
\label{eq:Ne23new}
&\frac{d\YY{23}{Ne}}{dt} = \YY{23}{Na} \lambda_7 - \YY{23}{Ne}
\lambda_8 - \YY{23}{Ne} \rho N_{\rm A} \bigl[\tilde Y(p)
\lambda_{11} + \tilde Y(\alpha) \lambda_{20}\bigr],\\
&\frac{d\YY{21}{Ne}}{dt} = \rho N_{\rm A} \biggl\lbrace
\YY{20}{Ne} \tilde Y(n) \lambda_{12} - \YY{21}{Ne} \bigl[ \tilde Y(p)
  \lambda_{18} + \tilde Y(n) \lambda_{16} + \tilde Y(\alpha) \lambda_{22}\bigr] \biggr\rbrace.
\end{align}

The energy generation rate can be obtained after a straightforward
modification of the individual terms above, noting that the combined
energy contribution of the inverse reaction
\reac{13}{N}{\gamma}{p}{12}{C} and the reaction
\reac{12}{C}{p}{\gamma}{13}{N} can be obtained simply by multiplying
the Q--value of reaction 3) to the associated term in the r.h.s. of
equation~(\ref{eq:C12new}).

\section{COMPARISON WITH THE DETAILED NETWORK}
In what follows we will compare the results of the full nuclear
network introduced in Section~\ref{sec:full} with those of the
simplified nuclear networks N1 and N2 introduced in
Sections~\ref{sec:1stapprox} and \ref{sec:2ndapprox}. We will compare
the evolution of the main and trace nuclei mole fractions, the time
evolution and the energy release in the form of photons or neutrinos
as a function of the fraction of burnt \xx{12}{C} nuclei.

In most examples we have chosen a temperature of $4\E{8}$ K and a
density of $3\E{9}$ g cm$^{-3}$, which are typical of the conditions
encountered during the thermonuclear runaway and before ignition in a
pre--supernova CO WD. We chose a temperature and density at which
the simplified solution errors would be representative of those
encountered at other temperatures and densities after the trace nuclei
have reached their equilibrium values. The time--scales for the trace
nuclei to reach their equilibrium values can be computed from
equations~(\ref{eq:approx_tau}).

\begin{figure}
  \centering
  \includegraphics[width=0.59\hsize,keepaspectratio]{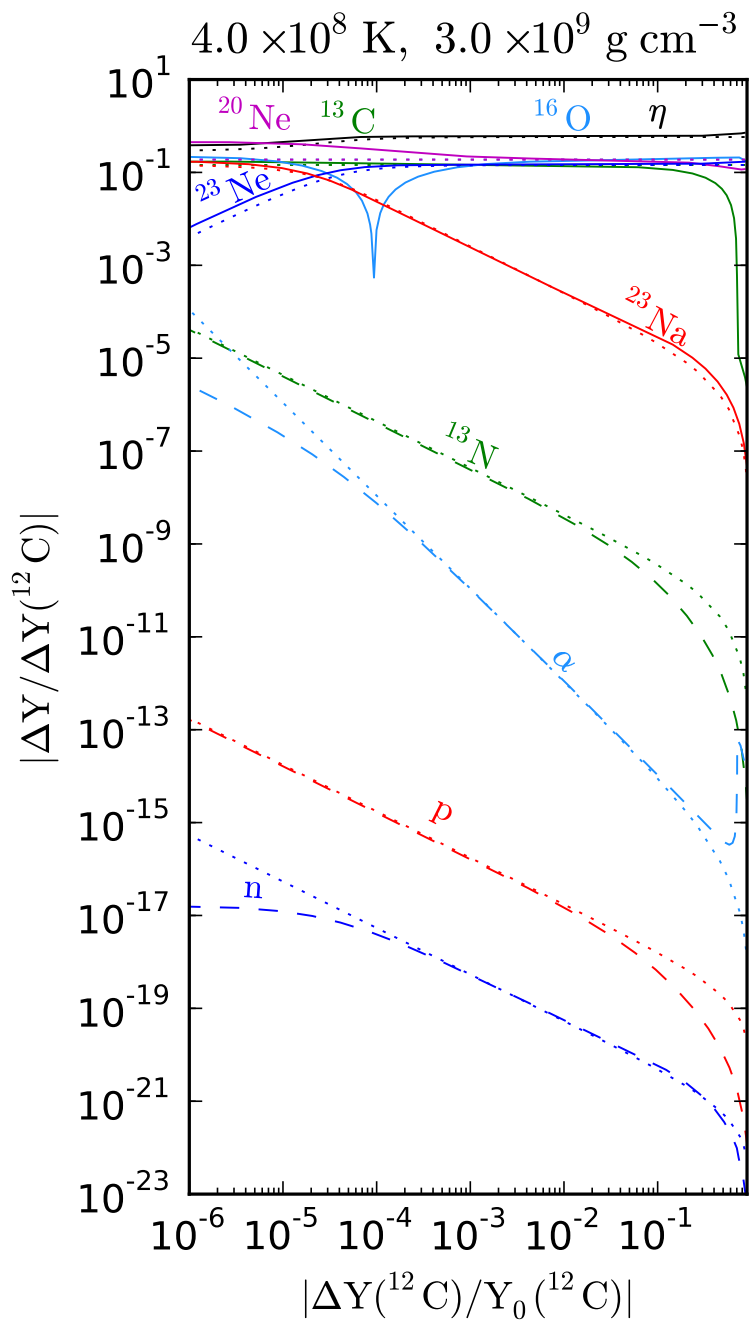}
  \caption[]{The ratio between the ashes and \xx{12}{C} mole fraction
    changes vs the fraction of burnt \xx{12}{C} in the full and N1
    networks. Main ashes and the neutron excess change are shown as
    continuous (full network) or dot--dashed (N1) lines and trace
    nuclei are shown as dashed (full network) or dotted (N1)
    lines. See discussion in the text.}
  \label{fig:Yevol_4e83e9_n_CO}
\end{figure}

\begin{figure}
  \centering
  \includegraphics[width=0.59\hsize,keepaspectratio]{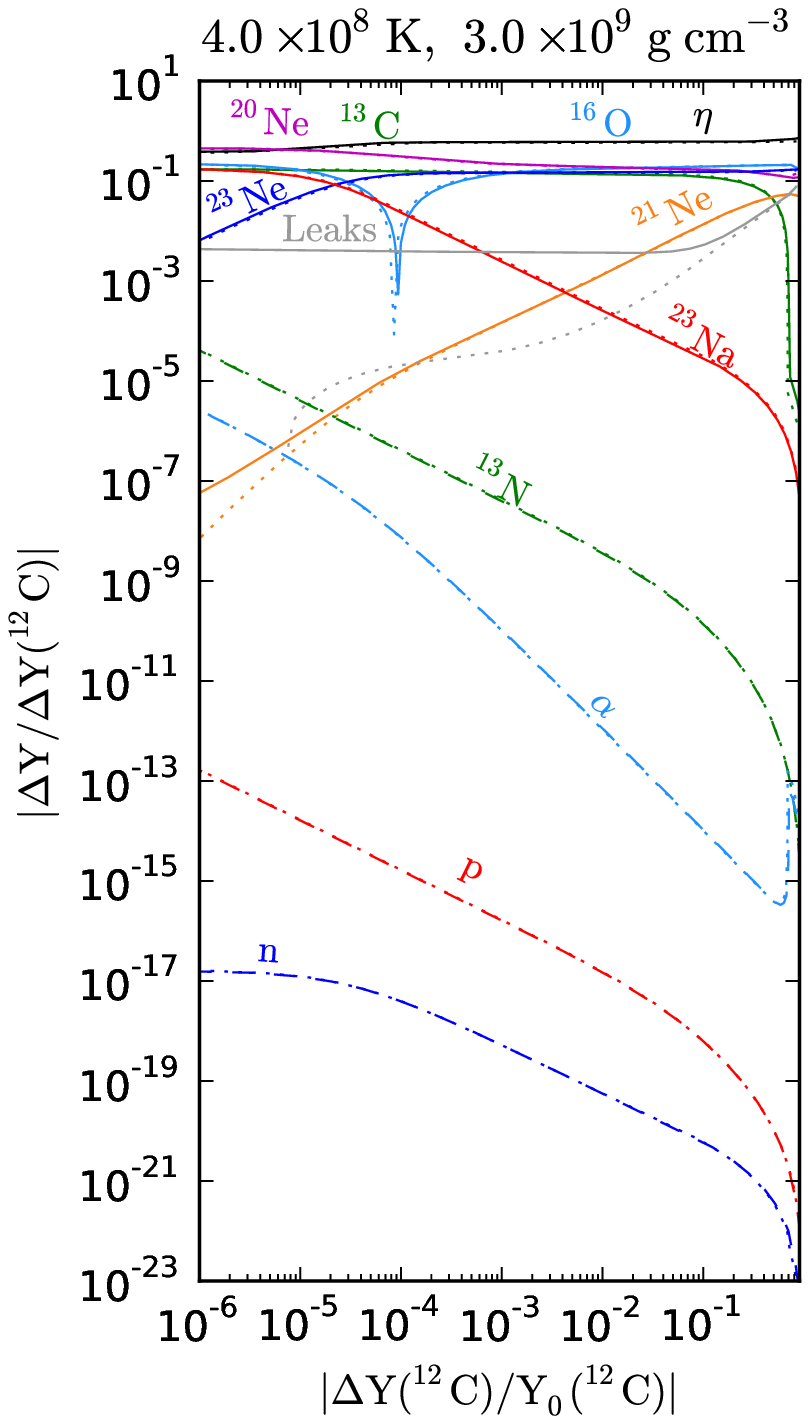}
  \caption[]{Same as Figure~\ref{fig:Yevol_4e83e9_n_CO}, but using the
    network N2 as comparison. We also include the mole fraction
    changes of \xx{21}{Ne} and the dummy leak particles introduced in
    equation~(\ref{eq:YL}). Note that the trace nuclei follow their
    equilibrium values precisely under this approximation. The
    abundance of leak nuclei is only well matched when more than
    $\approx$ 1\% of the \xx{12}{C} is burnt.}
  \label{fig:Yevol_4e83e9_n_CO_N2}
\end{figure}

In Figures~\ref{fig:Yevol_4e83e9_n_CO} and
\ref{fig:Yevol_4e83e9_n_CO_N2} we show the ratio between the
\xx{12}{C} ashes and \xx{12}{C} mole fraction changes vs the fraction
of burnt \xx{12}{C}, as well as the ratio between the neutron excess
change and the \xx{12}{C} mole fraction change vs the fraction of
burnt \xx{12}{C}. We compare the integration of the full network with
that of the networks N1 and N2, respectively. The transformation was
chosen because it makes the evolution of \xx{13}{C}, \xx{16}{O} and
\xx{20}{Ne} look exactly flat under N1, according to
equations~(\ref{eq:dC13C12}), (\ref{eq:dO16C12}) and
(\ref{eq:dNe20C12}), allowing an easier comparison between the
integrations of N1 and N2.

It can be seen that once the slowest evolving trace nuclei reach their
equilibrium value and before about 1\% of the \xx{12}{C} has been
burnt, both plots show a good agreement between the simplified and
full networks for the main \xx{12}{C} ashes. The exceptions are
$\alpha$--particles, \xx{16}{O} and \xx{20}{Ne} under N1 due to the
initial $\alpha$--leaks from reaction 15) and because we have assumed
a pure CO mixture, with \xx{13}{C} initially absent. Since in this
example the \xx{23}{Na} $e^-$--capture time--scale is shorter than the
\xx{23}{Ne} $\beta$--decay time--scale, both simplified solutions
correctly predict a higher abundance for \xx{23}{Ne} once both
time--scales become comparable. Both solutions show a good match for
the neutron excess changes too.

However, once the fraction of burnt \xx{12}{C} is above $\sim 1$\%,
the trace nuclei equilibrium abundances of N1 begin to differ from the
exact solution, whereas N2 matches their values even when more than
half of the original \xx{12}{C} has been burnt. The inclusion of the
leak reactions in N2 provides a better match for the trace nuclei and
as a consequence, the main \xx{12}{C} ashes as well,
once the trace nuclei equilibrium mole fractions have been
reached. Note also that the $\alpha$--particle, \xx{16}{O} and
\xx{20}{Ne} mole fraction changes are well matched in N2 once trace
nuclei equilibrium is reached.

In N2 the trace nuclei follow precisely their equilibrium values as
\xx{12}{C} burns, even if these equilibrium values change
dramatically. This is because their characteristic time--scales are
much smaller than the characteristic \xx{12}{C} burning
time--scale. This supports the use of N2 in a varying temperature and
density integration, which would be analogous to changing the
equilibrium values of the trace nuclei as \xx{12}{C} burns. The
shortest time--scale for the environment's variables to change should
be bigger than the biggest trace nuclei time--scale.

\begin{figure}
  \centering
  \includegraphics[width=\hsize,keepaspectratio]{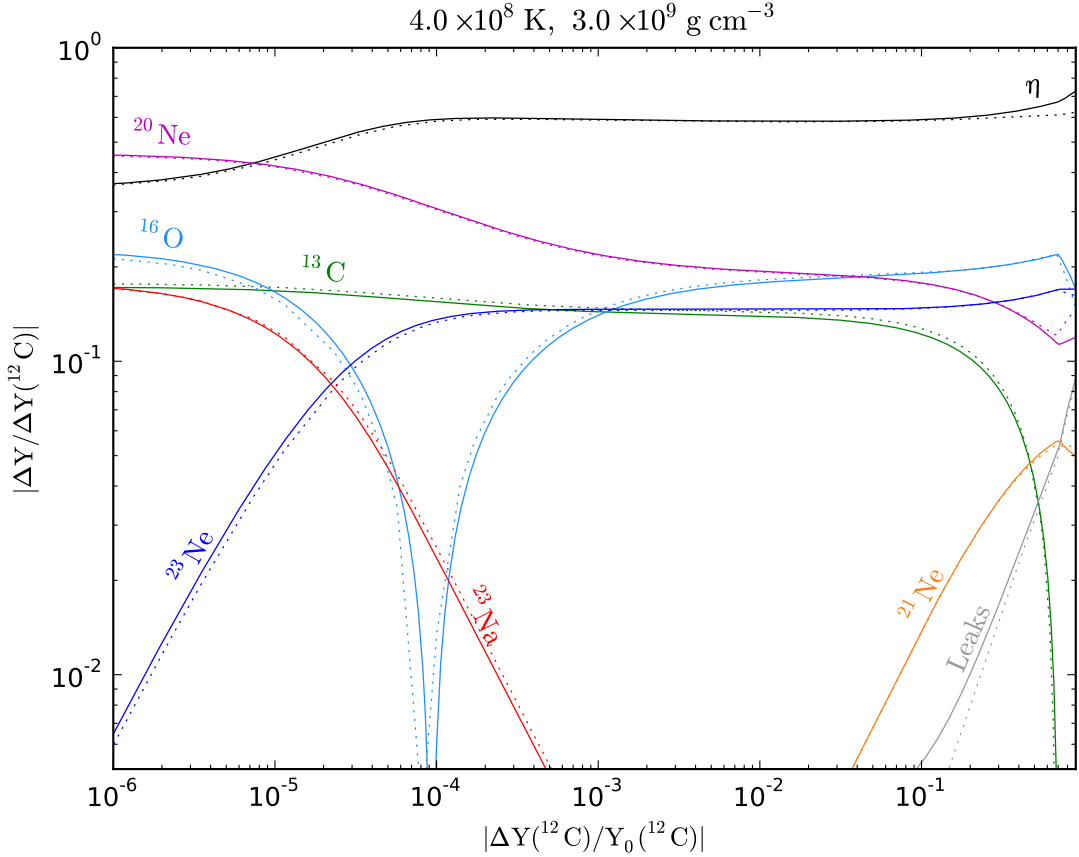}
  \caption[]{The ratio between the main ashes and \xx{12}{C} mole
    fraction changes vs the fraction of burnt \xx{12}{C} in the full
    (continuous) and N2 (dot--dashed) networks. Note that in N1 the
    evolution of the species \xx{13}{C}, \xx{16}{O} and \xx{20}{Ne}
    should be exactly flat in this plot according to
    equations~(\ref{eq:dC13C12}), (\ref{eq:dO16C12}) and
    (\ref{eq:dNe20C12}). Generally speaking, N2 reproduces the full
    network abundances to the 5\% level when the trace nuclei have
    reached their equilibrium mole fractions. In numerical
    experiments, N2 loses its accuracy when both $f_{\rm p}$ and
    $f_\alpha$ (see equations~\ref{eq:fp} and \ref{eq:fn}) are lower
    than 0.5, which is due to the observed \xx{13}{C} depletion.}
  \label{fig:Yevol_4e83e9_CO}
\end{figure}

In Figure~\ref{fig:Yevol_4e83e9_CO} we show an enlarged section of
Figure~\ref{fig:Yevol_4e83e9_n_CO_N2}. If N1 were valid, the evolution
of \xx{16}{O} and \xx{20}{Ne} would look flat and approximately 0.19
in this space. The same would be the case for \xx{13}{C}, although
approximately 0.15 (see equations~\ref{eq:dC13C12},
  \ref{eq:dO16C12} and \ref{eq:dNe20C12}). This is only
approximately true when the fraction of burnt \xx{12}{C} is between
2\% and 5\%. Since we plot absolute values of mole fraction changes,
Figure~\ref{fig:Yevol_4e83e9_CO} does not show whether the \xx{16}{O}
mole fraction is decreased or increased. In fact, its mole fraction is
originally depleted and, only after the fraction of burnt \xx{12}{C}
is about $10^{-4}$, it is increased. The original depletion is caused
by $\alpha$--leaks in reaction 15), which increase the mole fraction
of \xx{20}{Ne}.

The \xx{23}{Ne} mole fraction is larger than the \xx{23}{Ne} mole
fraction only after the burnt fraction of \xx{12}{C} is about
$2\E{-5}$. Interestingly, this makes the neutron excess evolution
change from above $0.3$ to $0.6$ in this transformation. This can be
understood noticing that $d\eta/dt = -2 dY_{\rm e}/dt$ and that the
value shown in equation~(\ref{eq:dYeC12}) is expected to double at
high density.

Note that \xx{13}{C} is depleted as \xx{21}{Ne} and leak nuclei
increase their abundance. This is because neutron leaks shortcut
\xx{13}{C} neutron captures, as discussed before. When this happens,
$f_\alpha$ decreases dramatically (see equation~\ref{eq:falpha}),
and the accuracy of the approximation is lost to about 20\%
level. Although at the beginning of the integration $f_\alpha$ is also
small, N2 matches the full network with great accuracy. To distinguish
between the cases when the accuracy of N2 is very good and only a
small fraction of \xx{12}{C} has been burnt from the late loss of
accuracy with a big fraction of \xx{12}{C} burnt, we define the
criterion for N2 to be considered an accurate representation of the
full network as
\begin{align} \label{eq:criterion}
  f_{\rm p} > 0.5 ~~~~{\rm and} ~~~~~f_\alpha > 0.5,
\end{align}
which should be used in a varying temperature and density integration,
such as in real stellar evolution models.

\begin{figure}
  \centering
  \includegraphics[width=\hsize,keepaspectratio]{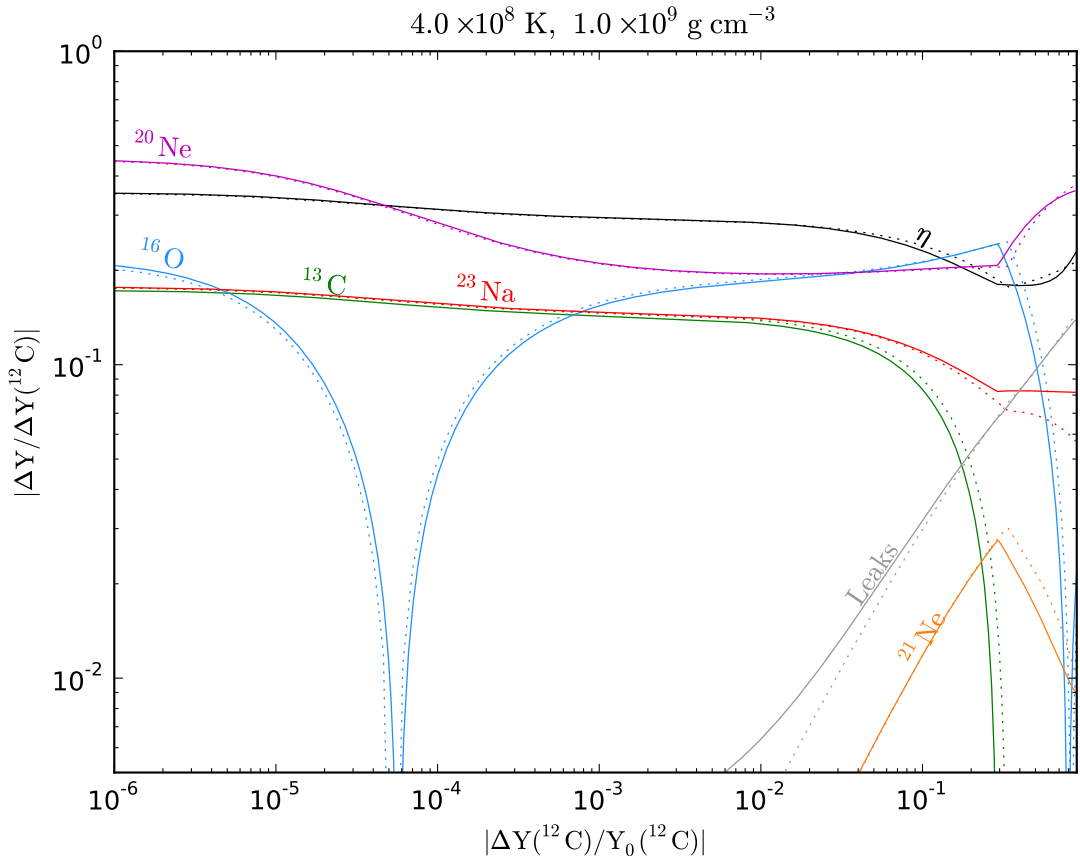}
  \caption[]{Same as Figure~\ref{fig:Yevol_4e81e9_CO}, but changing
    the density to $10^{9}$ g cm$^{-3}$. We can see that the network
    N2 loses its accuracy at a lower fraction of burnt \xx{12}{C},
    which is due to stronger neutron leaks which cause \xx{13}{C} to
    be relatively depleted sooner.}
  \label{fig:Yevol_4e81e9_CO}
\end{figure}

In Figure~\ref{fig:Yevol_4e81e9_CO} we show a similar integration as
in Figure~\ref{fig:Yevol_4e83e9_CO}, but at a density of $10^9$ g
cm$^{-3}$. We can see that \xx{13}{C} is depleted at a smaller
fraction of \xx{12}{C} than in Figure~\ref{fig:Yevol_4e83e9_CO}. This
is because there are more leak nuclei to capture neutrons. The same
associated loss of accuracy after \xx{13}{C} depletion is observed.
We can also see that the neutron excess evolution is closer to 0.3 in
this Figure, as expected from equation~(\ref{eq:dYeC12}). However, as
\xx{13}{C} is depleted, the e$^-$--capture rate decreases accordingly
and the neutron excess evolution becomes slower.

Finally, note that although by introducing the dummy leak nuclei we
can keep the number of independent variables small, we will
necessarily lose some accuracy matching the neutron excess. Many leak
nuclei can capture protons, neutrons or $\alpha$--particles, but they
can also capture electrons like \xx{23}{Na} does. Electron captures on
free protons were found to be negligible. In order of increasing
density, or Fermi electron energy, the following reactions may compete
with proton, neutron and $\alpha$--captures:
\ \reac{22}{Na}{e^-}{\nu_e}{22}{Ne},
\ \reac{24}{Na}{e^-}{\nu_e}{24}{Ne},
\ \reac{25}{Mg}{e^-}{\nu_e}{25}{Ne},
\ \reac{24}{Mg}{e^-}{\nu_e}{24}{Na}
\ or  \reac{21}{Ne}{e^-}{\nu_e}{21}{F}, affecting the neutron excess evolution \citep[see][]{1978ApJ...219..213I}. 

%

It is not possible to model the former electron captures accurately
without solving for each leak species and its electron capture
counterpart independently. Moreover, the relative abundances of the
different leak species will depend on the temperature and density
history of the gas, which makes it very difficult to compute the
contribution of the different electron capture reactions knowing only
$Y_{\rm L}$. However, assuming that the excess of neutrons over
protons of the leak nuclei is 1.5 under N2, a 10\% or 20\% accuracy
for the neutron excess evolution was achieved before and after
\xx{13}{C} depletion, respectively. This translates into a maximum of
5\% or 10\% error for the electron fraction, normally below 1\% for
small quantities of burnt \xx{12}{C}.

\subsubsection{Timing and Energy Generation Comparison}

\begin{figure}
  \centering
  \includegraphics[width=\hsize,keepaspectratio]{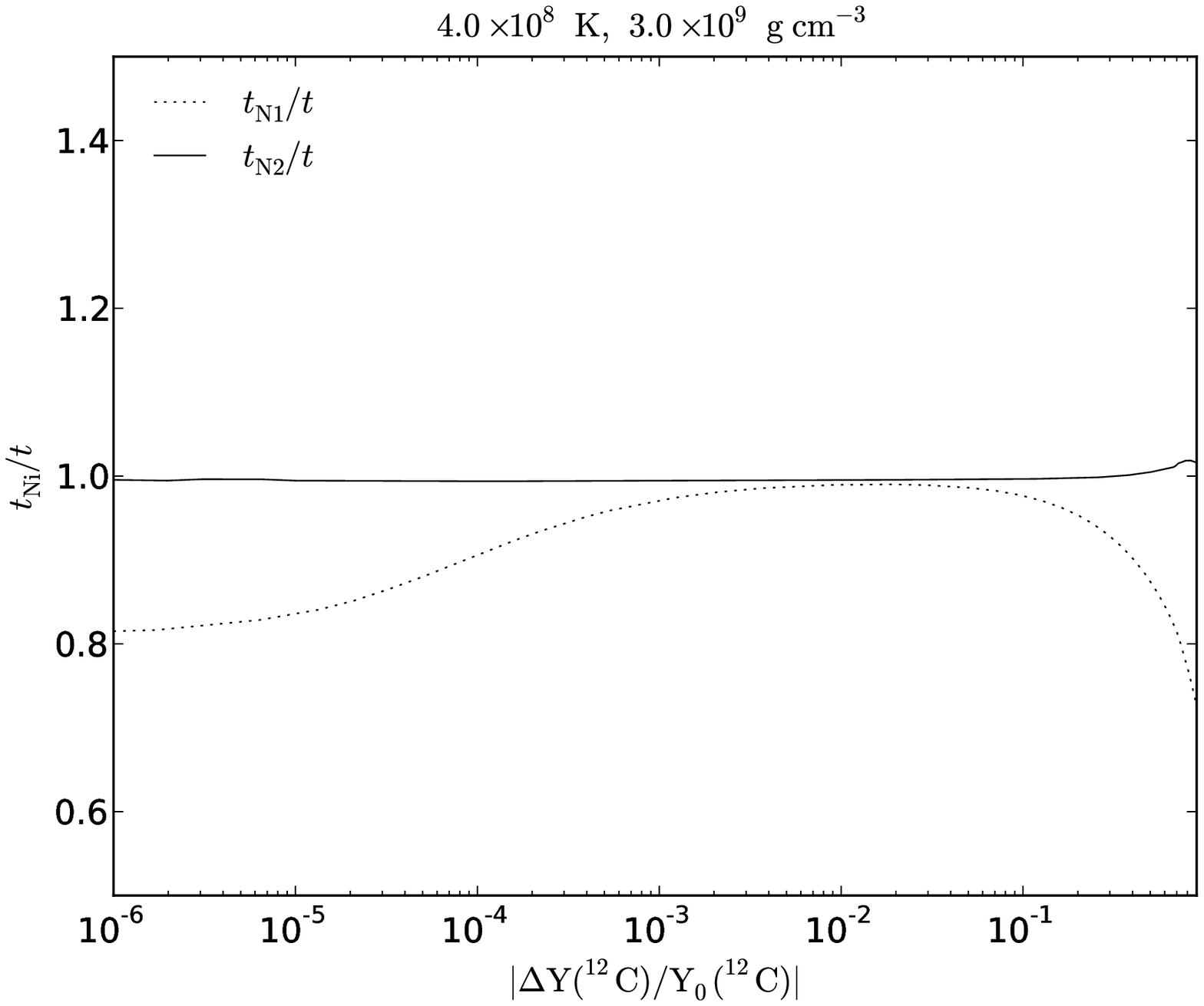}
  \caption[]{The ratio between the elapsed time in the two simplified
    networks and the full nuclear network plotted against the fraction
    of burnt \xx{12}{C}. N1 and N2 sub--scripts indicate which
    simplified network has been used for the comparison.}
  \label{fig:deltatime}
\end{figure}

In Figure~\ref{fig:deltatime} we show the ratio between the elapsed
times of the simplified and full integrations vs the fraction of burnt
\xx{12}{C}. The N1 and N2 sub--scripts correspond to the solution of
the first and second simplified networks, respectively. We can see
that the N1 can over--estimate the speed at which \xx{12}{C} burns by
more than 20\%, whereas the second simplified network reproduces the
time evolution to better than 5\% error. This is due to the absence of
leak reactions in the first approximation, which over--estimates the
amount of \xx{12}{C} proton and neutron captures at a given time.

\begin{figure}
  \centering
  \includegraphics[width=\hsize,keepaspectratio]{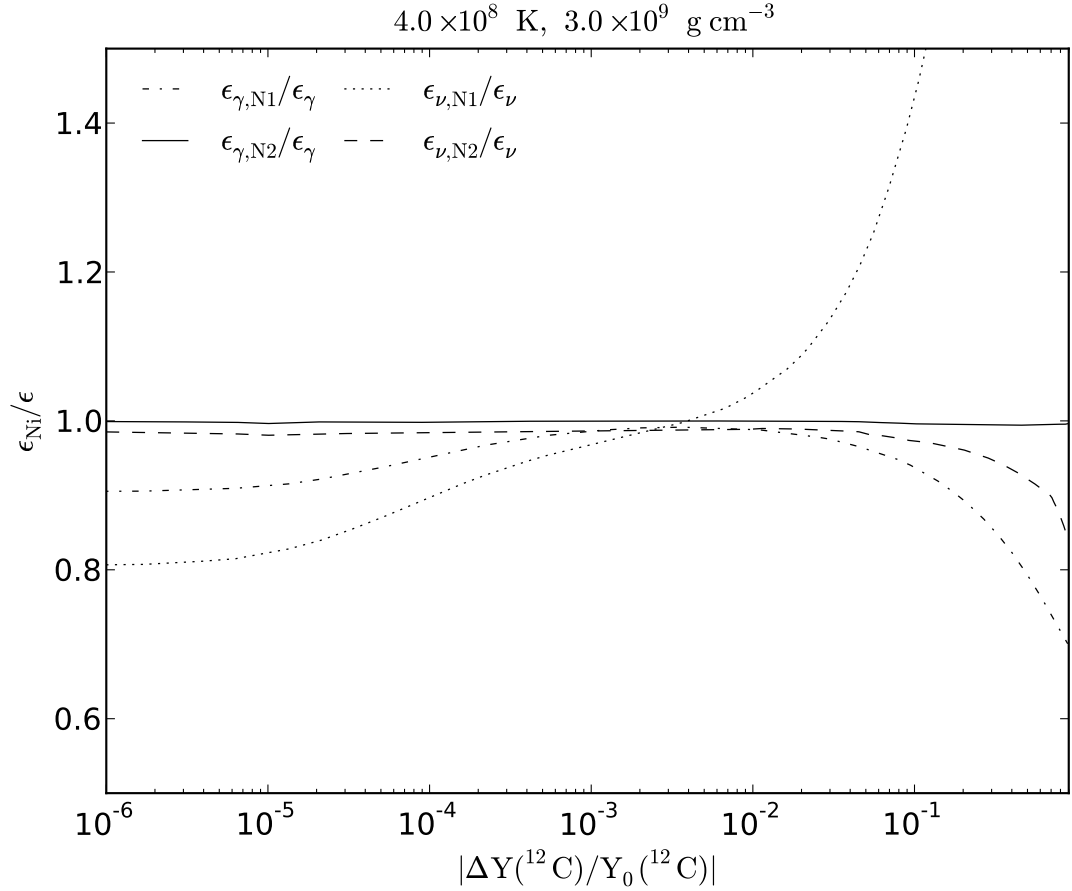}
  \caption[]{Same as Figure~\ref{fig:deltatime}, but we now show the
    ratios of the photon and neutrino energy release rates between the
    simplified and the full networks. N1 and N2 sub--scripts indicate
    which simplified network has been used for the comparison.}
  \label{fig:deltaenergy}
\end{figure}

In Figure~\ref{fig:deltaenergy} we show the ratios between the photon
and neutrino energy release rates in N1 and N2 and the photon and
neutrino energy release rate in the full network. We can see that the
photon energy release rates is typically off in N1 by 20\% or more,
whereas N2 matches them at the 5\% level, except after \xx{13}{C} is
depleted. This is because the leak reactions tend to release more
energy than the \xx{12}{C} proton and neutron captures. The neutrino
energy release rates are not as accurate because secondary electron
captures not included in these networks can have an important
contribution, but since photon energy rates are generally much bigger,
the error in the net energy release remains at the 5\% level.

\section{DISCUSSION AND CONCLUSIONS}
We have derived two approximate nuclear networks that can be used in
the hydrostatic carbon burning regime of carbon--oxygen white dwarfs
(CO WDs) approaching ignition. The networks have the advantage of
being able to track accurately the relevant species during this phase
of evolution without having to include the fast--evolving protons,
neutrons, $\alpha$--particles or \xx{13}{N} nuclei.

Using the same integration method \citep{baderdeuflhard83},
convergence tolerance and initial conditions in one--zone models, we
have found that the integration of N1 or N2 is much faster than that
of the detailed network.  Depending on the temperature, density and
composition, the ratio between the integration times of the detailed
network and N2 varied between a factor of a few and a factor of
several hundreds. Integration time ratios between N2 and N1 varied
between 1.2 and 1.3. Note that the slowest step in the integration of
full stellar evolution models is normally Gaussian elimination or
similar methods, which scale with the cube of the number of
independent physical variables. Thus, the simplified networks
presented in this work could have important applications in full
stellar evolution models.

Depending on the amount of \xx{12}{C} being burnt and the desired
errors in the mole fractions of the dominant nuclei, the time and the
net energy release, either the N1 or N2 approximations described in
Sections~\ref{sec:1stapprox} and \ref{sec:2ndapprox} could be used in
CO WD interiors. For an accuracy of the order of
50\%, and to have a simple picture of the main flows involved, N1 can
be used with caution. For an accuracy of the order of 5\%, N2 should
be used instead. If N2 is used and the criterion in
equation~(\ref{eq:criterion}) does not hold, only an accuracy of the
order of 20\% can be guaranteed, as long as all the trace nuclei
time--scales are much shorter than the \xx{12}{C}--burning
time--scale.

These accuracies are not valid for nuclei whose mole fractions are not
significant, for example \xx{23}{Ne} when the $\beta$--decay
time--scale is much shorter than the $e^-$--capture time--scale, since
in this case the \reac{22}{Ne}{n}{\gamma}{23}{Ne} reaction will
dominate its evolution, a reaction that cannot be included accurately
under N2. We have also found that for relatively low densities, below
$10^8$ g cm$^{-3}$, the quoted errors can double, but only when
significant amounts of \xx{12}{C} have been burnt, which normally
occurs when the density is significantly above this number. Thus, N2
is more accurate for central ignition models, but it could also be
used in off--center ignition models.

We have shown how to derive the approximations and have compared them
to a detailed network at fixed temperature and density. We have also
discussed how the approximations can break down and when they can be
used. Since all the details of the derivation are shown, these
simplified networks can be further improved straightforwardly. It is
also possible to build intermediate approximations between N1 and N2,
for example, including \xx{21}{Ne} in the dummy leak species defined
in equation~(\ref{eq:YL}) to remove one independent variable from the
solution. We have tested this last approximation in a few cases and
the resulting errors appears to be twice the errors in N2.

Although for simplicity we have only shown fixed temperature and
density integrations, these approximations can be used in environments
with varying temperature and density, such as real stellar evolution
models. This is because in the temperature and density range found in
pre--ignition WD interiors the trace nuclei reach their equilibrium
values much faster than the typical environmental variables vary
inside the star, even within strong pre--ignition convective velocity
fields.

The networks can account for \xx{23}{Na} or \xx{13}{N} $e^-$--captures
for the purposes of following the evolution of the
pressure--supporting electrons in pre--supernova WDs. We have shown
that they will be valid even in convective WD interiors. We recommend
the use of these networks when the mass fractions of the most abundant
species are relevant, namely \xx{12}{C}, \xx{16}{O}, \xx{13}{C},
\xx{20}{Ne}, \xx{23}{Ne} or \xx{23}{Na}, or the electron mole
fraction, $Y_{\rm e}$, or the energy generation rates. We do not
recommend its use if the abundances of other nuclei not included in
this discussion are being studied.

We have introduced an important tool to understand the effect of the
\emph{convective Urca process} on the ignition conditions of SNe
Ia. We foresee the application of these simplified networks or their
modification in detailed one or multi--dimensional stellar evolution
models trying to understand pre--supernova CO WDs or similar objects
\citep[see e.g.][]{2006MNRAS.368..187L, 2009ApJ...704..196Z,
  2010arXiv1001.2165I}.
\section*{ACKNOWLEDGMENTS}
We thank Edward Brown, David Chamulak, Stephen Justham and Anthony
Piro for helpful discussions. F.F. acknowledges partial support from
STFC, from CONICYT through projects FONDAP 15010003 and BASAL PFB-06,
from the GEMINI-CONICYT FUND 32070022 and from the Millennium Center
for Supernova Science through grant P06-045-F funded by ``Programa
Bicentenario de Ciencia y Tecnolog\'ia de CONICYT'' and ``Programa
Iniciativa Cient\'ifica Milenio de MIDEPLAN''.

\clearpage

\end{document}